\documentclass[11pt]{article}
\usepackage[utf8]{inputenc}
\usepackage{jheppub}

\usepackage{epsfig,latexsym,cancel,amssymb,amsmath}
\usepackage{graphicx,subfigure}
\usepackage{epstopdf}
\usepackage{hyperref, graphicx, slashed, xcolor, amsmath}
\usepackage{multirow}

\newcommand{\tr}[1]{\text{#1}}
\newcommand{\brr}[1]{\left(#1\right)}
\newcommand{\srr}[1]{\left[#1\right]}
\newcommand{\h}{{\rm h}}
\newcommand{\tp}{{\tilde{p}}}

\newcommand{\beq}{\begin{equation}}
\newcommand{\eeq}{\end{equation}}
\newcommand{\bea}{\begin{eqnarray}}
\newcommand{\eea}{\end{eqnarray}}
\newcommand{\tG}{{\tilde G}}
\newcommand{\cG}{{\tilde{\cal G}}}
\newcommand{\cF}{{\tilde{\cal F}}}
\newcommand{\cE}{{\tilde{\cal E}}}
\newcommand{\tomega}{\tilde{\omega}}
\newcommand{\tT}{{\tilde T}}
\newcommand{\hT}{{\hat T}}
\newcommand{\aend}{a_{\rm end}}
\newcommand{\He}{H_{\rm end}}

\newcommand{\tth}{{\tilde h}}
\newcommand{\hk}{h_{\bf k}}
\newcommand{\bk}{{\bf k}}
\newcommand{\bx}{{\bf x}}

\newcommand{\nn}{\nonumber \\}

\title{Gravitational Waves from an Inflation Triggered First-Order Phase Transition}

\author[a,b,c]{Haipeng An}
\author[d,e]{Kun-Feng Lyu}
\author[f,g]{Lian-Tao Wang}
\author[h]{Siyi Zhou}
\affiliation[a]{Department of Physics, Tsinghua University, Beijing 100084, China}
\affiliation[b]{Center for High Energy Physics, Tsinghua University, Beijing 100084, China}
\affiliation[c]{Center for High Energy Physics, Peking University, Beijing, 100871, China}
\affiliation[d]{School of Physics and Astronomy, University of Minnesota, Minneapolis, MN 55455, U.S.A.}
\affiliation[e]{Department of Physics, the Hong Kong University of Science and Technology, Clear Water Bay,
Kowloon, Hong Kong S.A.R., P.R.C.}
\affiliation[f]{Enrico Fermi Institute, University of Chicago, Chicago, IL 60637, USA}
\affiliation[g]{Kavli Institute for Cosmological Physics, University of Chicago, Chicago, IL 60637, USA}
\affiliation[h]{Department of Physics, Kobe University, Kobe 657-8501, Japan}

\emailAdd{anhp@mail.tsinghua.edu.cn}
\emailAdd{  lyu00145@umn.edu}
\emailAdd{  liantaow@uchicago.edu}
\emailAdd{  siyi@people.kobe-u.ac.jp}

\preprint{KOBE-COSMO-21-18}

\abstract{Large excursion of the inflaton field can trigger interesting dynamics. One important example is a first-order phase transition in a spectator sector which couples to the inflaton.  
Gravitational waves (GWs) from such a first-order phase transition during inflation, an example of an instantaneous source, have an oscillatory feature. In this work, we show that this feature is generic for a source in an era of accelerated expansion. We also demonstrate that the shape of the GW signal contains information about the evolution of the early universe following the phase transition. In particular, the slope of the infrared part of the GW spectrum is sensitive to the evolution of the Hubble parameter when the GW modes reenter the horizon after inflation. The slope of the profile of the intermediate oscillatory part and the ultraviolet part of the GW  spectrum depend on the evolution of the Hubble parameter when the modes exit horizon during the inflation and when they reenter the horizon during the reheating. 
The ultraviolet spectrum also depends on the details of the dynamics of the phase transition. We consider the GW signal in several models of evolution during and after inflation, and compare them with the minimal scenario of quasi-de Sitter inflation followed by radiation domination after a fast reheating, and demonstrate that the shape of the GW can be used to distinguish them.  In this way, the GW signal considered in this paper offers a powerful probe to the dynamics of the early universe which is otherwise difficult to explore directly through CMB, large scale structure, big bang nucleosynthesis (BBN), and other well-studied cosmological observables.
}

\begin{document}

\maketitle

\section{Introduction and summary}
\label{sec:intro}
It is highly plausible that the universe experienced an inflationary era before the hot big bang~\cite{Guth:1980zm,Linde:1981mu,Albrecht:1982wi}. The simplest models assume that the inflation is driven by the potential energy of a scalar field, sustaining an exponential expansion for about $40-60$ e-folds. While this picture could be too simplistic, for large classes of models, this is at least a good approximation for the epoch of inflation relevant to the cosmic microwave background measurements~\cite{Planck:2018jri}.

\begin{figure}[h!]
\centering
\includegraphics[height=3.0in]{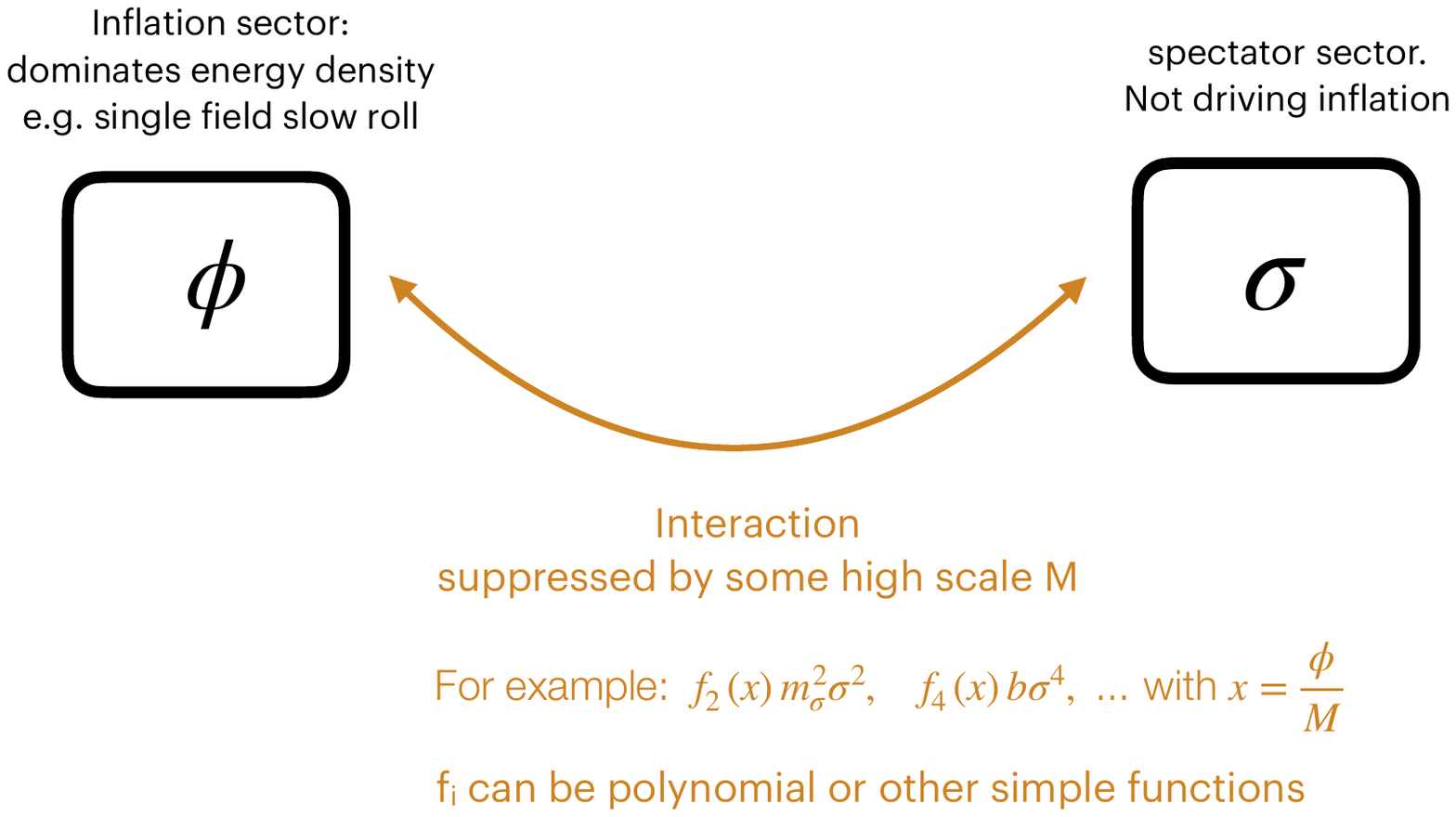}

\vspace{-0.7cm} 
\caption{Inflaton with coupling to an spectator sector. 
}\label{fig:InflatonSpectatorCoupling}
\end{figure}

At the same time, the dynamics during the inflation can be much richer than the picture of a single slowly rolling scalar field. We expect, generically, the inflaton should have couplings to other spectator fields. 
Due to the  relative flatness of the inflaton potential, we typically consider models where there is no significant coupling between the inflaton sector and the other spectator sectors. However, such coupling could still be induced by interactions at higher scales, as illustrated in Fig.~\ref{fig:InflatonSpectatorCoupling}. In this case, the strengths of the couplings can be much suppressed and do not affect the dynamics of the inflaton significantly. A typical example is a moduli field as  the inflaton. In general, the dynamics of a spectator sector are controlled by a set of terms of the form $\lambda_i {\mathcal{O}}_i$, where ${\mathcal{O}}_i$ is an operator consisting of spectator fields, and $\lambda_i$ is its coupling. The inflaton controls the size the couplings $\lambda_i$ through 
\begin{equation}
\lambda_i (\phi) = \lambda_i^{0} f_i \left( \frac{\phi}{M}\right),
\end{equation}
where $f_i(x)$ is a dimensionless function, and $M$ is a fundamental scale characterizing the strength of the coupling between the inflaton and the spectator sector. An obvious example would be $M=M_{\rm Pl}$, as in the case of the inflaton being a string theory moduli field.  When the inflaton field eventually settled down to its minimum $\phi_{0}$, the low energy coupling of the spectator fields will be $\lambda_i (\phi_0)$. 

At the same time, during the inflation, the inflaton field can have excursions comparable to $M_{\rm Pl}$. In fact, the excursion of the inflaton field can be estimated as
\begin{equation}
\Delta \phi \sim N_{{\rm e}-{\rm fold}} \sqrt{\epsilon} M_{\rm Pl}~.
\end{equation}
Hence, for the class of models in which the slow roll parameter $\epsilon$ is not vanishingly small, the inflaton can travel a large distance during inflation even during a small number of e-folds.
This can result in ${\mathcal{O}} (1)$ variations of coupling $\lambda_i$, which can trigger significant changes in the dynamics of the spectator sector. Of course, the high scale may not necessarily be the Planck scale, it could also be string scale, GUT scale. Another example is that, if the parameters of the spectator sector are determined supersymmetry breaking,  the high scale in the coupling is the messenger scale.

It has been conjectured that a field excursion greater than $M_{\rm Pl}$ cannot come from a consistent theory of quantum gravity \cite{Ooguri:2006in}. The signal we are interested in here does not rely on the validity of this conjecture.  Even in the case of $M\simeq M_{\rm Pl}$, it is enough to have the field excursion {\it comparable} to $M_{\rm Pl}$ (instead of being parametrically larger) to trigger the phase transition we discuss here. 

There are many possible scenarios with significant changes in the dynamics of the spectator sector triggered by the excursion of the inflaton field. Perhaps one of the most dramatic possibilities would be a first-order phase transition. As a toy model, we could consider the case of a spectator scalar field $\sigma$ with coupling to the inflaton of the form\footnote{It would be clear from the subsequent discussion that much of the conclusions of this paper does not depend on the details of this model. 
 We will leave a fuller exploration of the model space for a future work.  Possible scenarios are also studied in Ref.~\cite{Jiang:2015qor}, where first-order phase transition in the thermal plasma is assumed to present at the beginning of inflation, and in Ref.~\cite{Sugimura:2011tk}, where the inflation starts from a first-order phase transition. And in~\cite{Ashoorioon:2015hya,Ashoorioon:2020hln,Ashoorioon:2022raz}, such kind of first-order phase transition happens at the end of inflation. Some results about the features of the GW spectrum from first-order phase transition are also studied in Ref.~\cite{Wang:2018caj}. } 
\begin{equation}
	V(\phi, \sigma) =- \frac{1}{2}\left(\mu^2 - c^2 \phi^2  \right) \sigma^2+ \dots~,
\end{equation}
where $c$ is a dimensionless numerical coefficient. During inflation, the sign of the mass square of $\sigma$ flips, which can trigger a phase transition. A detailed study of the phase transition dynamics is provided in appendix~\ref{model}. There are a couple of general requirements we impose on the spectator sector. First of all, by definition, we would consider the inflaton sector dominates the energy density. At the same time, we would also focus on the case in which the phase transition is strongly first-order. In addition, we would like to be in a scenario in which the true vacuum would take   up $O(1)$ fraction of the Hubble volume at the end of the phase transition. 
As shown in Ref.~\cite{An:2020fff}, these requirements can be satisfied with 
\begin{equation}
\beta^4 \ll m_\sigma^4 \ll 3 M_{\rm Pl}^2 H^2_\star, 
\label{eq:scalerequirement}
\end{equation}
where $\beta^{-1} \equiv r_{\rm bubble}$ is the size of the typical bubble at the completion of the phase transition. $H_\star$ is the Hubble at the phase transition $\tau = \tau_\star$. 
For the class of models we consider, we find 
\begin{equation}
r_{\rm bubble} \equiv \beta^{-1} \sim \left(10^{-2} - 10^{-1}\right) \times H^{-1}_\star.  
\end{equation}
Hence, at the completion of the phase transition, a typical Hubble volume will be filled with many smaller bubbles of the true vacuum\footnote{In principle, a phase transition in the spectator sector can back-react on the inflaton and influence its dynamics. In general, close to the critical point, the degrees of freedom in the spectator sector can become light and this effect can become non-negligible. This is interesting for further study. In this paper, we focus on the scenario in which the phase transition happens during the epoch which can not be probed by CMB and large scale structure observations. Toegther with the assumption that the spectator sector is subdominant in energy density, we do not expect a strong constraint from such a back-reaction.}.

\begin{figure}[h!]
\centering
\includegraphics[height=2.5in]{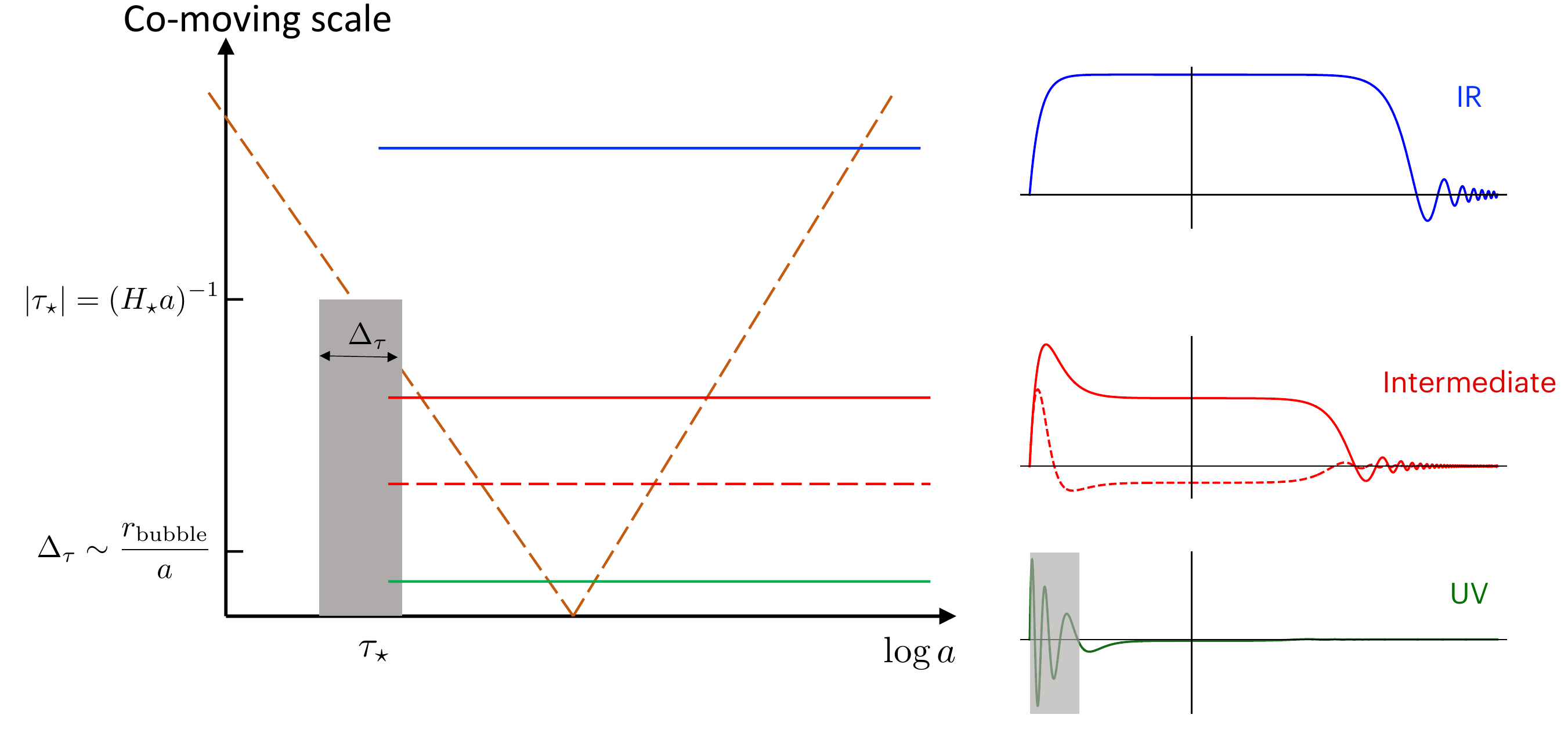}
\caption{Illustration of various relevant scales and corresponding GW signal. The bubbles nucleated during a phase transition collide, which give rises to a GW source around $\tau_\star$ lasting for a short duration of $\Delta_\tau \ll (H_\star a_\star)^{-1}$, where $H_\star$ and $a_\star$ are the Hubble and the scale factor at the phase transition, respectively. }\label{fig:schematic}
\end{figure}
At the end of the phase transition, the collisions of the bubbles generate GWs. The focus of our paper is on the properties of this GW signal, and using it as a probe to the expansion history of the early universe. The duration of the bubble collisions, $\Delta_\tau$ can be estimated to be the same order as the size of the bubbles right before the completion of the phase transition
\begin{equation}
\Delta_\tau \sim \frac{r_{\rm bubble}}{a_\star}  \ll (H_\star a_\star)^{-1}~,
\end{equation}
where $a_\star \equiv a(\tau=\tau_\star)$ is the scale factor at the phase transition. 
Since the duration of the bubble collision is much shorter than the Hubble time during the phase transition, it can be considered as an approximately {\it instantaneous} source of the GW. The relevant scales in our discussion are shown in Fig.~\ref{fig:schematic}.  The signal produced in this way will be a stochastic GW background, and can be detected by terrestrial or space GW telescopes~\cite{Seoane:2013qna,Audley:2017drz,Kawamura:2011zz,Luo:2015ght,Guo:2018npi,Crowder:2005nr,Harry:2006fi,Corbin:2005ny,Kramer:2013kea,Hobbs:2009yy,Janssen:2014dka,TheLIGOScientific:2014jea,Abramovici:1992ah,TheVirgo:2014hva,Punturo:2010zz,Reitze:2019iox}.

\begin{figure}[h!]
\centering
\includegraphics[height=2.5in]{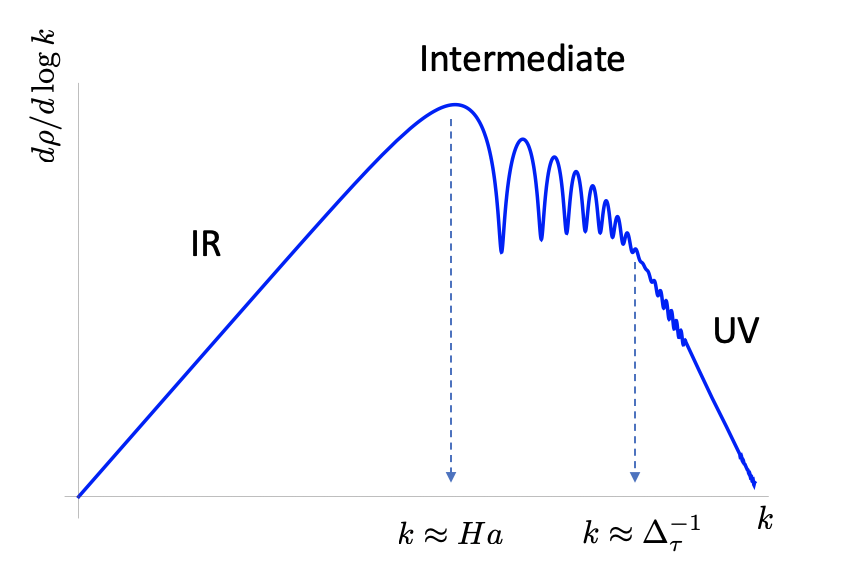}
\caption{Illustration of the $k$ (frequency) dependence of different parts of a typical GW spectrum. The signal typically peaks around $k \approx H_\star a_\star$, with an oscillatory feature in the range $H_\star a_\star < k < \Delta^{-1}_{\tau}$. }\label{fig:genericS}
\end{figure}

The spectrum of the signal depends on its frequency~\cite{An:2020fff}, or equivalently, the corresponding co-moving momentum $k$. Qualitatively, we can consider the  three different frequency ranges, IR, intermediate, and UV, which we present as blue, red and green curves in Fig.~\ref{fig:schematic}. An example of the spectrum of the GW signal is shown in Fig.~\ref{fig:genericS}.  More specifically, 
\begin{enumerate}
\item[${\bullet }$] IR: $k<H_\star a_\star$ (blue curve in Fig.~\ref{fig:schematic}). The modes in this range are outside of the horizon when they are produced during the bubble collision $\tau \sim \tau_\star$. They will not oscillate until evolving back into the horizon after reheating. The slope of the IR part of the GW spectrum can be written as $[\cE^i_0(k)]^2 k^5$. $\cE^i_0(k)$, which we will discuss in detail later,  describes the evolution of the universe when the modes evolve back into the horizon. 
\item[${\bullet }$] Intermediate (oscillatory): $H_\star a_\star < k < \Delta_\tau^{-1}$ (red curves in Fig.~\ref{fig:schematic}). In this range, the modes oscillate after being generated at $\tau \sim \tau_\star$ until they exit the horizon. This gives rise to an oscillatory pattern in frequency in the GW signal, which will be smeared by the finite duration of the bubble collision $\Delta_\tau$.  In addition to the evolution of the universe, when the GW mode reenters the horizon, it is also sensitive to the evolution of the universe before the GW mode leaves the horizon, parameterized by a factor $\cG^f_0(k)^2$. The slope of the profile of the oscillatory part can be written as $[\cE^i_0(k) \cG^f_0(k)]^2 k^3$. Moreover, the slope of this part is independent of the details of bubble collisions. 
\item[${\bullet }$] UV: $k> \Delta_\tau^{-1}$ (green curve in Fig.~\ref{fig:schematic}). The GW signal in this range does not show an oscillatory pattern, which is smeared out completely due to the finite duration of the source. The slope of the UV part of the GW spectrum can be written as $|\hat T_{ij}(k, \bk)|^2 [\cE^i_0(k) \cG^f_0(k)]^2 k^3$.  The factor  $|\hat T_{ij}(k, \bk)|^2$ is the energy-momentum tensor of the source. For GW generated by a first order phase transition, it depends on the details of the dynamics of the bubble collisions. Therefore, at least in principle, with the knowledge of $\cE^i_0(k)$ and $ \cG^f_0(k)$ from the measurement of the the IR and the oscillatory parts  of the signal,  the UV spectrum can be used to determine the detailed mechanism of the GW production. 
\end{enumerate}

Perhaps the simplest picture of the early universe contains a single period of quasi-de Sitter (dS) inflation. At the end of the inflation, there is a  quick reheating followed by  radiation dominated (RD) expansion. However, the actual evolution can be much more complicated.  The inflationary era does not have to be quasi-de Sitter. In addition, the inflation can have  different stages, some of them could be close to quasi-de Sitter but others not. During the reheating, if the conversion from the energy in the inflaton to the radiation is not very efficient, the universe will be matter dominated (MD) for a significant period of time. In the RD period, if there are some long lived matter, it is possible that they will dominate the energy density, leading to stages of MD. As the universe is cooling down, there can also be phase transitions, which can produce other topologically defects (such as cosmic strings) temporarily dominating the energy density of the Universe. If some of these new dynamics happen during the CMB or large scale structure modes exit or re-enter the horizon, or during some later epoch such as the big bang nucleosynthesis (BBN), there could be corresponding observational signals. Otherwise, if these would happen in between these epochs, we would have few direct probes. For example, even in the simplest scenario with inflation driven by a slow rolling scalar field, which is also responsible for reheating, we still have little handle on the details of the inflaton potential for the last ten(s) e-foldings before reheating. 

\begin{figure}[h!]
\centering
\includegraphics[height=2.2in]{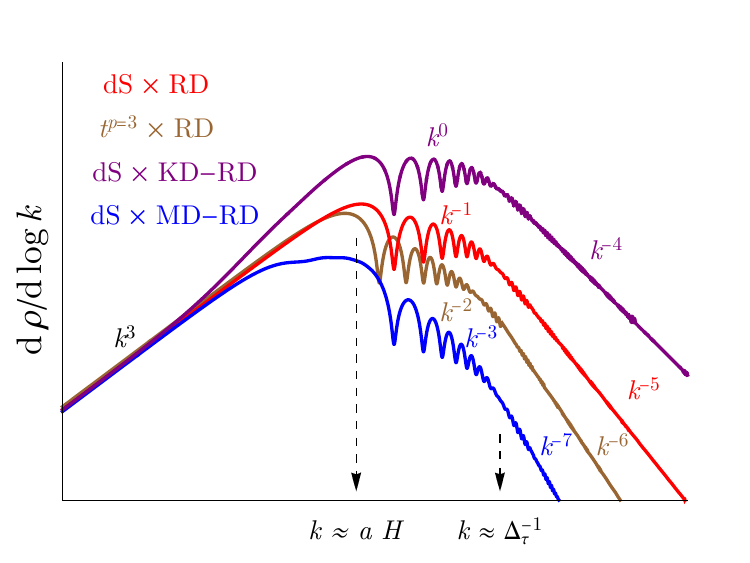}
\includegraphics[height=2in]{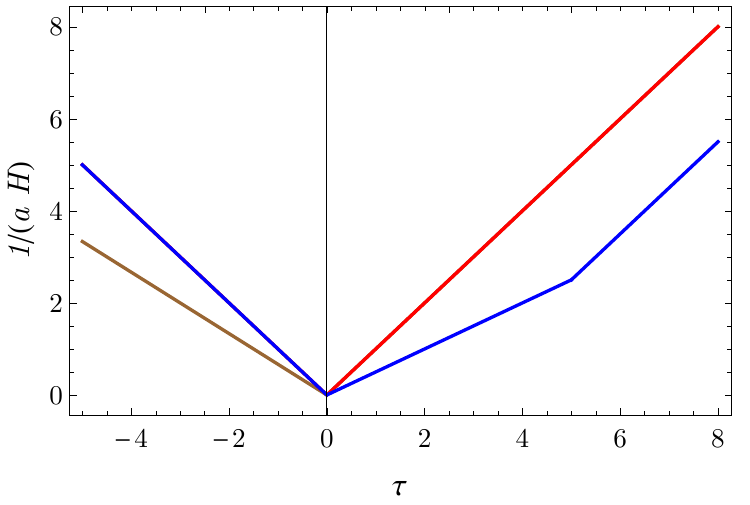}
\caption{On the left panel, we compare spectral shapes of the GW signal in a few non-minimal scenarios of the early universe, including an intermediate period of matter domination (blue) or a kination domination (purple) after the end of inflation,  and a $t^p$-inflation scenario with $p=3$ (brown). On the right panel, we plot the co-moving horizon as a function of conformal time for these scenarios. We fix the scale factor $a$ and the conformal time $\tau_\star$ when phase transition take place. 
The same rules apply for the comoving horizon plots in the rest of the paper.}\label{fig:scenarios}
\end{figure}

From the discussion above, it is obvious that the GW signal discussed in this paper offers an opportunity in filling in a big gap we have in probing the history of the early universe. As we will demonstrate in detail in this paper, the shape of the GW signal, encapsulated in factors $\cG^f_0(k)$ and $\cE^i_0 (k)$, depends sensitively on the evolution of the universe. In particular, we consider the GW signal in a set of alternative models of the evolution of the early universe,  and compare them with the minimal scenario. An example of such a comparison is shown in  Fig.~\ref{fig:scenarios}, which includes scenarios with an intermediate period of MD  and kination domination (KD) after the end of inflation,  and a $t^p$-inflation scenario with $p=3$. The features of the GW spectrum in these scenarios will be derived in detail later in this paper. At the same time, it is clear from Fig.~\ref{fig:scenarios} that the spectral shape of the GW signal can distinguish these different scenarios. 
If GWs with the oscillatory feature described in this paper is observed, it would offer an unmistakable signal for an approximate instantaneous source in the early universe. At the same time, it would be of great interest to measure its shape in detail, which would be powerful in distinguishing different scenarios of the history of the early universe~\cite{Hook:2020phx}.

We emphasize that the form the gravitation wave signal stems mainly from the existence of an approximately {\it instantaneous} source of GW during the inflation.
We focus here on the case of first-order phase transition triggered by a rolling inflaton as a plausible candidate of such a source. Another possible source could be a local feature on the inflaton potential \cite{Fumagalli:2021mpc}. At the same time,  stochastic GW background can be produced by other sources as well. For example, it can be produced through 
black hole/neutron star mergers~\cite{Mandic:2016lcn,Clesse:2016ajp,Wang:2016ana,Raidal:2017mfl,Garcia-Bellido:2017aan,Guo:2017njn}, preheating~\cite{Khlebnikov:1997di,Easther:2006vd,GarciaBellido:2007dg,GarciaBellido:2007af,Dufaux:2007pt}, decay of cosmic strings~\cite{Vachaspati:1984gt,Brandenberger:1986xn,Hindmarsh:1990xi,Damour:2001bk,Siemens:2001dx,Hindmarsh:1994re}, and primordial magnetic field~\cite{Durrer:1999bk,Caprini:2001nb,Pogosian:2001np,Caprini:2003vc,Caprini:2006jb,Caprini:2009pr,Shaw:2009nf,Saga:2018ont}.
It can also be produced through quantum fluctuation during inflation~\cite{Grishchuk:1974ny,Starobinsky:1979ty,Rubakov:1982df,Fabbri:1983us,Abbott:1984fp}. Large stochastic GW with peak structures can also be produced in single field inflation model with inflection point~\cite{Ballesteros:2020qam,Bhaumik:2019tvl,Bhaumik:2020dor,Ragavendra:2020sop}, non-canonical kinetic terms of the inflaton field~\cite{Lin:2020goi,Yi:2020cut,Zhang:2020uek}, axion inflation with explosive production of gauge fields~\cite{Cook:2011hg,Barnaby:2011qe,Namba:2015gja,Garcia-Bellido:2016dkw,Ozsoy:2020ccy,Ozsoy:2020kat}. Multiple peaks in the GW spectrum can be produced in models with periodic inflaton potentials, axion inflation~\cite{Ozsoy:2020ccy}, primordial magnetic field~\cite{Caprini:2006jb,Caprini:2009pr}, peaks on the primordial power spectrum~\cite{Witkowski:2021raz,Fumagalli:2021dtd,Fumagalli:2020nvq,Braglia:2020taf} and GWs from density perturbations in an early MD era~\cite{Dalianis:2020gup}. However, none of these scenarios can produce features in the GW spectrum similar to first-order phase transitions during inflation.

The rest of the paper is organized as the following. The main features of the GW signal have been worked out in Ref.~\cite{An:2020fff}. In Sec.~\ref{sec:generic}, we review the key arguments and give more detailed derivations of the spectral shape of the GW signal. 
In Sec.~\ref{sec:alternative}, we discuss how to use the oscillatory pattern to distinguish different inflation models. In Sec.~\ref{sec:later}, we discuss the influence on the oscillatory pattern from later evolutions of the universe after inflation. In Sec.~\ref{sec:compare}, we compare the qualitative features of the GW spectrum for different scenarios. To be complete, in Sec.~\ref{sec:signal}, we compare the GW signal to the sensitivities of the planned GW observatories.   
We summarize our results and discuss future directions in Sec.~\ref{sec:summary}.

\section{Features of the GW signal}
\label{sec:generic}

In this section, we give a detailed derivation of the spectral shape of the GW signal. 
The metric of our expanding universe can be written as
\bea
ds^2 &=& - d t^2 + a^2(t)(\delta_{ij} + h_{ij})dx^i dx^j \ , \nonumber \\
& =& a^2(\tau) \left[-d\tau^2 + (\delta_{ij} + h_{ij})dx^i dx^j \right] ~,
\eea
where $a$ is the scale factor, $\tau$ denotes the conformal time with $d\tau = a^{-1}(t) d t$, and the transverse traceless part of $h_{ij}$ parameterizes the GW degrees of freedom. 
During accelerated expansion of the universe, there is an event horizon which plays a key role in shaping the GW spectrum. 
The GW perturbation (in the following of the paper, we use $h_{ij}$ to describe only its transverse and traceless part) satisfies the differential equation 
\bea\label{eq:diff}
h''_{ij} + \frac{2 a'}{a} h'_{ij} - \nabla^2 h_{ij} = 16\pi G_N a^2 \sigma_{ij} \ ,
\eea
where $'$ indicates derivatives with respect to the conformal time $\tau$, $G_N$ is the Newton's gravity constant, and $\sigma_{ij}$ is the transverse, traceless part of the energy-momentum tensor. For convenience, we define 
\bea
\h_{ij} (\tau,\bx) = a(\tau) h_{ij} (\tau, \bx) \ ,
\eea
which then satisfies 
\bea\label{eq:hh}
\h_{ij}'' - \left(\nabla^2 + \frac{a''}{a}\right) \h_{ij} = 16 \pi G_N a^3 \sigma_{ij} \ .
\eea
The solution of Eq.~(\ref{eq:hh}) can be written as
\bea
\h_{ij}(\tau, \bx) = \int d\bx' d\tau' G_R(\tau, \tau';\bx - \bx') 16\pi G_N a^3(\tau') \sigma_{ij}(\tau',\bx') \ ,
\eea
where the retarded Green's function $G_R$ satisfies
\bea
\left[\frac{\partial^2}{\partial\tau^2} - \nabla^2 - \frac{a''}{a}\right]G_R(\tau, \tau'; \bx - \bx')  = \delta(\tau-\tau')\delta^3(\bx-\bx') \ .
\eea
The Fourier transformation of $G_R$ satisfies
\bea\label{eq:green}
\tG''_R (\tau,\tau';\bk) + \left(k^2 - \frac{a''}{a}\right) \tG_R(\tau,\tau';\bk) = \delta(\tau - \tau') \ .
\eea
We introduce a new dimensionless variable $\eta=k \tau$ and  a new Green's function 
\bea
\cG(\eta, \eta') = k \tilde G_R(\tau, \tau' ; \bk) \ ,
\eea
which satisfies
\bea\label{eq:cgreen}
\left(\frac{d^2}{d\eta^2} + 1 - \frac{d^2 a}{a d\eta^2}\right) \cG(\eta,\eta') =  0  \ ,
\eea
with the initial condition
\bea\label{eq:initial1aaa}
\cG(\eta',\eta') = 0 \ ,\;\; \left.\frac{d\cG(\eta,\eta')}{d\eta}\right|_{\eta=\eta'} = 1\ .
\eea
For the convenience of the later discussion, we also introduce another solution of Eq.~(\ref{eq:cgreen}), ${\cF}(\eta, \eta')$, satisfying initial condition
\bea\label{eq:initial2aaa}
\cF(\eta',\eta') = 1 \ ,\;\; \left.\frac{d\cF(\eta,\eta')}{d\eta}\right|_{\eta=\eta'} = 0\ .
\eea

\subsection{$H_\star a (\tau_\star)< k < \Delta_\tau^{-1}$, oscillatory pattern in the GW signal}
\label{sec:2point1}

In this sub-section, we offer a detailed derivation of the presence of the oscillatory feature in the intermediate range of co-moving wave number $H_\star a (\tau_\star)< k < \Delta_\tau^{-1}$.
\subsubsection{Oscillatory pattern of the Green's function}
\label{subsec:green}

We begin with a discussion of the evolution of the GW before the end of the inflation. During the accelerated expansion epoch, the {\it effective mass} term in Eq.~\ref{eq:cgreen}, $a^{-1}d^2 a/ d\eta^2 = (2k^2)^{-1} d^2 a^2/dt^2$, is always positive. 
 For the modes we are interested in, the propagation of the GW during the inflation era naturally 
  separates into three periods. A mode will exit horizon at $\eta = \eta_{\rm exit}$.  In the region $\eta \ll \eta_{\rm exit} \simeq -1$, the mode is deeply inside the horizon and the curvature of the spacetime becomes unimportant, so that we can neglect the  effective mass term in Eq.~(\ref{eq:cgreen}). In the region $|\eta| \gg \eta_{\rm exit} $ the mode is outside the horizon and we can neglect the ``1'' term in Eq.~(\ref{eq:cgreen}). For $\eta \sim {\mathcal O} (\eta_{\rm exit})$ , we cannot ignore either of the two contributions, and analytic solutions can be found only in special cases. 

Towards the end of inflation, the modes are outside of the horizon with $\eta \to 0$. Hence, the behavior of $\cG/a$ as $\eta \to 0$ determines the shape of GW signal before reheating. The key observation is that, for generic inflation models,  $\lim_{\eta\rightarrow0}\cG/a$ can always be written as
\bea\label{eq:hGf}
\left.\frac{\cG(\eta,\eta')}{a(\eta)}\right|_{\eta\rightarrow0} = \cos(\eta'-\eta_0) \cG_0^f \ ,
\eea
where $\eta_0$ is a $k-$dependent phase. Both $\eta_0$ and $\cG_0^f$ depend on the details of the inflation models. In particular, the dependence of $\cG_0^f$ on $k$ can tell us the information of the evolution of $a$ before the modes with comoving momentum $k$ exit the horizon. 

To understand  Eq.~(\ref{eq:hGf}), we begin by considering the Green's function with a reference point  $\bar \eta$, at which the corresponding mode is still deep inside the horizon. We can write, for $\eta > \bar \eta$,  
\bea \label{eq:combo}
\cG (\eta, \eta') = c^{(1)} \cG (\eta, \bar\eta ) + c^{(2)} \cF (\eta, \bar\eta ). 
\eea
In the limit of $\eta, \ \eta' \ll \eta_{\rm exit}$, $\cG (\eta, \eta') \to \sin (\eta - \eta')$ and $\cF (\eta, \eta') \to \cos (\eta - \eta')$.  Taking this limit on the both sides of  Eq.~(\ref{eq:combo}), we conclude
\bea
c^{(1)} = \cos (\bar \eta - \eta')~, \ c^{(2)} = \sin (\bar \eta - \eta')~. 
\eea
Hence, we have 
\bea\label{endofinf}
\left.\frac{\cG (\eta,\eta')}{a(\eta)}\right|_{\eta\rightarrow0} = \cos(\bar \eta - \eta') \ \left.\frac{\cG(\eta, \bar \eta ) }{a(\eta)}\right|_{\eta\rightarrow0} + \sin(\bar \eta- \eta') \ \left.\frac{\cF(\eta,\bar \eta)}{a(\eta)}\right|_{\eta\rightarrow0} \ ~.
\eea
We can always rewrite Eq.~(\ref{endofinf}) in the form of Eq.~(\ref{eq:hGf}), with an additional phase factor $\eta_0$.

In summary, what we have shown is that once $\eta'$ is in the range of $\eta' \ll 1$, the value of $\cG/a$ far outside the horizon has a cosine dependence on $\eta'$ as given in Eq.~(\ref{eq:hGf}). The value of $\cG^f_0$ is independent of the choice $\eta'$. 
The asymptotic value, $\cG^f_0$ depends on $k$, as well as the scale factor $a$.
Therefore, $\cG_0^f$ is sensitive to the detailed evolution of the universe when the modes exit the horizon. Later in this paper, we will demonstrate how to use it as a tool to probe different models of early universe.

\subsubsection{Oscillatory pattern in GW spectrum}
\label{subsec:osc}

Before the end of the inflation, using Eq.~(\ref{eq:hGf}),  the GW signal outside the horizon can be written as
\bea\label{eq:hf}
\tth^f_{ij}(\bk) = \frac{16\pi G_N  \cG_0^f}{k}  \int d\tau'  \tT_{ij}(\tau',{\bf k})  \ \cos[k(\tau' - \tau_0)] \ ,
\eea
where $\tau_0 = \eta_0 / k$, and $\tth^f_{ij}$ is a Fourier mode  of the $h_{ij}$. $ \tilde T_{ij}(\tau',\bk)$ is  {physical} energy-momentum tensor in Fourier space, 
\bea
\tilde T_{ij}(\tau',\bk) = a^{3}(\tau') \int d^3 x e^{- i \bk \cdot {\bf x}} \sigma_{ij}(\tau , \bx)~. 
\eea

After inflation, the evolution of the $\tth_{ij}(\tau, \bk)$ satisfies 
\bea\label{eq:diffhk}
\tth''_{ij}(\tau, \bk) + \frac{2 a'}{a} \tth'_{ij} (\tau, \bk) + k^2 \tth_{ij}(\tau, \bk) = 0 \ ,
\eea
which can also be written as 
\bea
\frac{d^2}{d \eta^2}(a \tth_{ij} ) + \left(1 - \frac{d^2 a}{ad \eta^2}\right) (a \tth_{ij}) = 0 \ . 
\eea
Since the modes we are interested in are already outside the horizon right before the end of the inflation, we set the initial condition of $\hk$ as 
\bea \label{init2}
\tth_{ij}(0,\bk) = \tth_{ij}^f \ , \;\; \tth'_{ij}(0, \bk) = 0 \ ,
\eea
where $\tth^f(\bk)$ is given in Eq.~(\ref{eq:hf}). 
The general solution of $\hk$ can be written as
\bea
\tth_{ij}(\tau,\bk) = \tth^f_{ij} (\bk) {\cal E}(k\tau) \ ,
\eea
where ${\cal E}$ satisfies
\bea\label{eq:Einitial}
{\cal E} (0) = 1 \ ,\;\; {\cal E}' (0) = 0 \ .
\eea

After the mode evolves back into the horizon ($k \gg \dot a = a'/a$),
\bea\label{eq:cEi}
{\cal E}(\eta) = {\cE}^i_0 a^{-1} \sin(\eta + \phi) \ .
\eea
Here the coefficient of proportionality $\cE^i_0$ also depends on $k$, just like $\cG^f_0$, and from now on we will write its dependence on $k$ explicitly.

The observed energy density at conformal time $\tau$ can be written as
\bea
\rho_{\rm GW} = \frac{1}{16\pi G_N a^2} \langle h'^2_{ij}(\tau,\bx) \rangle \ ,
\eea
where $\langle\cdots \rangle$ denotes the the spatial average and the average over $\tau$ for at least several periods to measure the GW.  We have 
\bea
\langle h'^2_{ij}(\tau,\bx) \rangle = \frac{1}{V} \int \frac{d^3k}{(2\pi)^3} \frac{d^3k'}{(2\pi)^3}\int d^3x \tth'_{ij} (\tau,\bk) \tth'^*_{ij}(\tau,\bk') e^{i(\bk-\bk')\cdot \bx} = \frac{1}{V}\int\frac{d^3k}{(2\pi)^3} \langle |\tth'_{ij}(\tau,\bk)|^2 \rangle \ ,\nn
\eea
where $V$ is the total comoving spatial  volume. Hence, 
\bea\label{eq:rhoGW}
\rho_{\rm GW}(\tau) &=& \int \frac{d^3k}{(2\pi)^3} \frac{8\pi G_N\left[\cE^i_0(k) \cG^f_0(k)\right]^2}{V a^4(\tau)} \nn & & \times \int d\tau_1' d\tau_2' \cos k(\tau_1' - \tau_0) \cos k(\tau_2'-\tau_0)~ \tT_{ij} (\tau_1',\bk_p) \tT^*_{ij} (\tau_2',\bk_p) \ ,
\eea
where the average over $\tau$ gives $\sin^2k\tau \rightarrow 1/2$. 

The GWs are produced by bubble collisions for a duration of  $\Delta_\tau$ around $\tau_\star$. 
We can expand the cosine factor around $\tau' = \tau_\star$,
\bea
\cos k (\tau'-\tau_0) = \cos (k(\tau_\star - \tau_0) + k \Delta_
\tau) = \cos k(\tau_\star-\tau_0)\cos k\Delta_\tau-  \sin k(\tau_\star-\tau_0) \sin k \Delta_\tau \ . \nn
\eea
For $k \Delta_\tau \ll 1$, at zeroth order,
$\cos k \Delta_\tau \rightarrow 1 \ , \;\; \sin k \Delta_\tau  \rightarrow 0 \ .$
Carrying out the $\tau_1'$ and $\tau_2'$ integrals, we have
\bea
\rho_{\rm GW}(\tau) = \int \frac{d^3k}{(2\pi)^3} \frac{8\pi G_N\left[\cE^i_0(k) \cG^f_0(k)\right]^2}{V a^4(\tau) a^2(\tau_\star)} \cos^2 k(\tau_\star - \tau_0') \hT_{ij}(0,\bk) \hT_{ij}^*(0,\bk) \ . 
\eea       
where $\hT_{ij}(0,\bk) = a(\tau_\star)\int d \tau \tT_{ij}(\tau,\bk) $ 
is the zero mode of the temporal Fourier transformation of $\tT_{ij} (\tau,\bk)$. 
As shown in Ref.~\cite{Cai:2019cdl}, and also numerically in Ref.~\cite{Huber:2008hg} for a large class of models including first-order phase transition, in the case that the physical momentum, $\bk_p \equiv \bk/a$, is smaller than all the energy scales in the GW source, $\langle \hT_{ij}(0,\bk) \hT_{ij}^*(0,\bk)\rangle$ is independent of $k_p$. Here the $\langle \cdots \rangle$ should be understood as a statistical average. 
Thus, the differential spectrum can be written as
\bea\label{eq:drho}
\frac{d\rho_{\rm GW}}{d\log k} = \frac{4 G_N |\hT_{ij} (0, 0)|^2 }{\pi V a^4(\tau) a^2(\tau_\star)}  \left\{ \left[\cE^i_0(k) \cG^f_0(k)\right]^2 k^3 \cos^2 k(\tau_\star - \tau_0) \right\} \ .
\eea
This is the general formula for GW originated from instantaneous sources during inflation. The $k$ dependences are collected in the factor within $\{\cdots\}$. The factors $\cE^i_0(k)$ and $\cG^f_0(k)$ and $\tau_0$ are model dependent. The oscillatory pattern in the power spectrum is explicitly shown by the $\cos^2 k(\tau_\star - \tau_0)$ factor. 

\subsubsection{The smearing effect from the finite duration of the sources}
\label{subsec:smearing}

In practice, the duration of the GW source, $\Delta_\tau$ is finite. 
If GW is from incoherent sources, such as bubble collisions, the finite duration is expected to smear the oscillation. To study this quantitatively, let's impose the condition that $| k \tau_\star | \gg 1$. This condition guarantees that the space-time can be treated as flat during the production of GW. 
The $\tau_1'$ and $\tau_2'$ integrals in Eq.~(\ref{eq:rhoGW}) can be written as 
\bea\label{eq:248}
&&\int d\tau_{1}' d\tau_{2}' \left[c_* \cos k (\tau'_1 - \tau_\star) - s_* \sin  k (\tau'_1 - \tau_\star)\right]  \left[c_* \cos k (\tau'_2 - \tau_\star)  - s_* \sin k (\tau'_2 - \tau_\star) \right] \nn
&&\quad \times \tT_{ij}(\tau_1',\bk_p)  \tT^{*}_{ij}(\tau_2',\bk_p) \nn
&=& \frac{1}{4a^2(\tau_\star)} \left[ |\hT(k, \bk)|^2 + |\hT(-k,\bk)|^2  
+ 2 \cos 2k(\tau_\star - \tau_0)~{\rm Re}(\hT(k,\bk)\hT^*(-k,\bk))\right. \nn &&\left. + 2 \sin 2k(\tau_\star - \tau_0)~{\rm Im}(\hT(k,\bk)\hT^*(-k,\bk))\right] \ . 
\eea
In the limit that $k \rightarrow 0$, we have 
\bea
\langle  |\hT(k, \bk)|^2 \rangle \ , \;\;  \langle  |\hT( - k, \bk)|^2 \rangle \ , \;\; {\rm Re}\langle(\hT(k,\bk)\hT^*(-k,\bk))\rangle\;\; \rightarrow \;\; \langle  |\hT(0, 0)|^2 \rangle \ ,
\eea
and we reproduce the result in Eq.~(\ref{eq:drho}). 

Eq.~(\ref{eq:248}) shows that  the size of $\langle(\hT(k,\bk)\hT^*(-k,\bk))\rangle$ determines the amplitude of the oscillation. Now, let's study how this term varies with the relative sizes of $k^{-1}$ and the duration of the source. We know that the energy-momentum tensor in coordinate space must be real. Therefore, we have $\hT^*(-k,\bk) = \hT(k,-\bk)$. Hence, $\langle(\hT(k,\bk)\hT^*(-k,\bk))\rangle = \langle(\hT(k,\bk)\hT(k,-\bk))\rangle$ is the correlation of $\hT$ at opposite directions. Therefore, this correlation is expected to vanish once $k$ is much larger than the typical scale or $\Delta_\tau^{-1}$. Since we assume $k|\tau_\star| \gg 1$, we can discuss this in flat spacetime. In general, we have
\bea
\langle(\hT(k,\bk)\hT^*(-k,\bk))\rangle = \int d\tau_1 d\tau_2 e^{i k (\tau_1 + \tau_2)}\tT (\tau_1, \bk) \tT^* (\tau_2, \bk) \ .
\eea
Decomposing $\tau_1$ and $\tau_2$ into $\tau_c\equiv (\tau_1+\tau_2)/2$ and $\delta \tau = (\tau_2 - \tau_1)$, the above integral becomes
\bea
\int d\tau_c ~ e^{i k \tau_c}  \left[\int d\delta \tau ~\tT (\tau_c-\delta \tau, \bk) \tT^* (\tau_c+\delta \tau, \bk) \right] \ ,
\eea
where the factor in the $[\cdots]$ only depends on the property of the source. Therefore, if $k > \Delta_\tau^{-1}$ the factor $e^{i k\tau_c}$ will fast oscillate and suppress the $\tau_c$ integral. Moreover, for $k < \Delta_\tau^{-1}$, there can still be a smearing effect. 
For example, if the shape of the source as a function of time is Gaussian like, then the smearing factor is 
$e^{ - k^2 \Delta_\tau^2/2}$. If it is square like, the smearing factor becomes $\sin(k \Delta_\tau) / (k \Delta_\tau)$. For a realistic model, the smearing factor can be determined from numerical simulation (see~\cite{Huber:2008hg} for results using envelope approximation and~\cite{Cutting:2018tjt,Gould:2019qek} for results away from envelope approximation). In this paper, we choose the simple smearing factor, namely assuming the GW signals at different conformal time are incoherent so eventually we just averaged the squared amplitude with time.

As a result, the smeared oscillatory part of the spectrum can be generally estimated as 
\bea\label{eq:smeared}
\frac{d\rho_{\rm GW}^{\rm osc}}{d\log k} = \frac{2 G_N |\tT_{ij} (0, 0)|^2 }{\pi V a^4(\tau) a^2(\tau_\star)}  \left\{ \left[\cE^i_0(k) \cG^f_0(k)\right]^2 k^3  \left[1 + {\cal S}(k\Delta_\tau)\cos2k(\tau_\star-\tau_0)\right]\right\} \ ,
\eea
where ${\cal S}$ is the smearing factor. 

Although similar oscillations can also show up in models where GW is generated by primordial magnetic fields, those are due to coherent superposition~\cite{Caprini:2009yp,Caprini:2009fx,Caprini:2007xq}. Incoherent superposition of post inflationary sources cannot generate oscillating feature on the GW spectrum.

\subsection{$k> \Delta_\tau^{-1}$, the UV behavior of the GW spectrum} 
\label{sec:UV}

In this case, the discussions in the Sec.~\ref{subsec:smearing} still applies. Therefore, the contribution from the oscillating term in Eq.~(\ref{eq:248}) is suppressed. Therefore, in this region, we have
\bea\label{eq:UV2}
\frac{d\rho_{\rm GW}^{\rm UV}}{d\log k} = \frac{2 G_N |\hT_{ij} (k, \bk)|^2 }{\pi V a^4(\tau) a^2(\tau_\star)}  \left\{ \left[\cE^i_0(k) \cG^f_0(k)\right]^2 k^3  \right\} \ .
\eea
The UV spectrum depends on both the detailed models of inflation, later evolution after inflation, and the information of the GW source. 
The details of the GW signal at the source can be obtained by numerical simulations of the bubble collision. Since the size of the bubble is much smaller than the curvature of the spacetime during the bubble collision, the numerical simulation carried out in flat space still applies to our case. The details of our numerical treatment will be discussed in Sec.~\ref{sec:general_form}.

In the limit of flat space-time, the total energy density of GW spectrum induced by a general GW source can be written as~\cite{Weinberg:1972kfs}
\bea\label{eq:GWflat}
\frac{d \rho^{\rm flat}_{\rm GW}}{d\log k_p} = \frac{2 G_N}{\pi V_{\rm phy}} k_p^3 \hat T^*_{ij}(k, \bk) \hat T_{ij} (k, \bk) \ ,
\eea
where $V_{\rm phy}$ is the total physical volume of the space. $k_p$ is the physical momentum $k_p \equiv k/a$. 
Comparing the GW spectrum in flat spacetime to the UV part of the GW spectrum in Eq.~(\ref{eq:UV2}),  we have 
\bea\label{eq:UV}
\frac{d\rho^{\rm UV}_{\rm GW}}{d\log k} = \frac{d\rho^{\rm flat}_{\rm GW}}{d \log k_p} \left[ \cE^i_0(k) \cG^f_0(k)\right]^2 \left(\frac{a(\tau_\star)}{a(\tau)}\right)^4 \ ,
\eea
where $\tau$ is the time when the GW is observed.

\subsection{$k< H_\star a(\tau_\star)$, the IR behavior of GW spectrum}\label{sec:IRGW}

In this regime, the GW mode is already outside of the horizon when it is produced by the phase transition. 
In this case, the Green's function ~(\ref{eq:cgreen}) can be simplified as 
\bea
\left( \frac{d^2}{d\eta^2} - \frac{d^2 a}{a d\eta^2} \right) \cG(\eta, \eta') = 0 \ ,
\eea
with the initial condition 
\bea
\cG(\eta',\eta') = 0\ ,\;\;\; \left.\frac{d\cG(\eta,\eta')}{d\eta} \right|_{\eta=\eta'} = 1\ .
\eea
The solution to $\cG$ at $\eta\rightarrow0$ can be written as
\bea
\cG^f_0(k) = \left.\frac{\cG(\eta,\eta')}{a(\eta)}\right|_{\eta\rightarrow0} = a(\eta') \int_{\eta'}^0 a^{-2}(\eta_1) d\eta_1 = k \left[a(\tau') \int_{\tau'}^0 a^{-2}(\tau_1) d\tau_1\right] \ ,
\eea
where in the last step the integral variable is changed back to $\tau$ and we can see that the factor in $[\cdots]$ is independent of $k$. Compared to the steps in Sec.~\ref{subsec:osc}, the infrared spectrum can be written as
\bea\label{eq:IR}
\frac{d\rho_{\rm GW}^{\rm IR}}{d\log k} &=& \frac{4 G_N |\hT_{ij}(0,0)|^2}{\pi^2 V a^4(\tau)} \left\{\left[ \cE^i_0(k) \cG^f_0(k) \right]^2 k^3 \right\}  \nn
&=& \frac{4 G_N |\hT_{ij}(0,0)|^2}{\pi^2 V a^4(\tau)} \left[ \int_{\tau_\star}^0 a^{-2}(\tau_1) d\tau_1 \right]^2 \left\{\left[ \cE^i_0(k) \right]^2 k^5 \right\} \ .
\eea
One can see that the infrared spectrum is fixed up to the model dependence of the evolution of the universe after inflation. 

Notice that in the IR regime, similar to the intermediate regime, both theoretical analysis and numerical simulation show that $ \hat T^*_{ij}(k, \bk) \hat T_{ij} (k, \bk)$ becomes independent of $k$ and approaches to $|\hat T_{ij}(0,0)|^2$, as we have done in Eq.~(\ref{eq:IR}). 

\subsection{General form of the GW spectrum }
\label{sec:general_form}

In this work, we focus on inflationary models in which the plasma energy is negligible compared to the vacuum energy. Furthermore, the typical radius of the bubbles at the completion of the phase transition is parametrically smaller than $H_{\star}^{-1}$. Hence, in calculating the energy-momentum tensor $\hat T^*_{ij} \hat T_{ij}$ of the source of the GW, we can simply neglect the expansion of the universe. 
This allows us to use the result of $\hat T^*_{ij} \hat T_{ij}$  produced by first-order phase transition in the RD era. The numerical calculations of the GW spectrum in the RD era has been done with the envelope approximation~\cite{Kosowsky:1992vn,Kosowsky:1992rz,Weir:2016tov,Huber:2008hg} as well as the lattice simulation~\cite{Cutting:2020nla,Cutting:2018tjt}. There are also analytical results for FLRW universe~\cite{Zhong:2021hgo}. More recently, there are also some work on improvements of envelope approximation~\cite{Konstandin:2017sat,Ellis:2020nnr,Lewicki:2020jiv,Lewicki:2020azd}. The results show that the GW spectrum produced by vacuum phase transition can be parameterized by a broken power law, 
\bea\label{eq:flat1p}
\frac{d\rho^{\rm flat}_{\rm GW}}{\Delta\rho_{\rm vac} d\log k_p} = \frac{\Delta \rho_{\rm vac}}{\rho_{\rm inf \star}}  \left( \frac{H_{\star}}{\beta} \right)^2\times \tilde\Delta \times \frac{(a+b) \tilde k_p^b k_p^a}{b \tilde k_p^{a+b} + a k_p^{a+b}} \ ,
\eea
where $\Delta\rho_{\rm vac}$ is the change of the vacuum energy between the true and false vacuum.
In the case of vacuum phase transition with the plasma effect neglected, the wall velocity approaches the speed of light. Different simulations gave somewhat different  values of $\tilde\Delta$. Using the envelope approximation, Ref.~\cite{Huber:2008hg} gives $\tilde\Delta\approx 0.077$ in the ultra-relativistic region, whereas a more recent work~\cite{Konstandin:2017sat} gives $\tilde\Delta\approx 0.47$. In the lattice simulation in Ref.~\cite{Cutting:2018tjt}, a more complicated expression in the form of $(a+b)^c \tilde k_p^b k_p^a / (b \tilde k_p^{(a+b)/c} + a  k_p^{(a+b)/c})^c$ is introduced to the fit the GW spectrum. The height of the peak from the lattice simulation is smaller than from the envelope approximation but within the same orders of magnitude. The main purpose of this work is to demonstrate the effect of the inflating universe on the GW spectrum. To this end, we fix  $\tilde\Delta = 0.077$ in our numerical results in the next sections. In Eq.~(\ref{eq:flat1p}), the peak wavenumber $\tilde k_p$ is determined by the size of the typical size of the bubble when the phase transition completes. In Refs.~\cite{Huber:2008hg, Konstandin:2017sat}, the value of $\tilde k_p$ is $2.16\beta$ and $1.96\beta$, respectively. In the following, we will use $\tilde k_p = 2\beta$. 

In the flat space-time limit, the IR and UV power law behaviors are determined by the indices $a$ and $b$ in Eq.~(\ref{eq:flat1p}), respectively. Both theoretical calculation~\cite{Cai:2019cdl} and causality argument~\cite{Caprini:2009fx} show that in the absence of long range interactions as in the case of first-order phase transition, $a = 3$ in the deep IR region. This is also confirmed by numerical simulations~\cite{Huber:2008hg,Konstandin:2017sat,Cutting:2018tjt}. For the slope of the UV part of the GW spectrum, the simulations using the envelope approximation shows $b \approx 1$~\cite{Huber:2008hg,Konstandin:2017sat}, while the lattice simulation shows $b = 1.5$~\cite{Cutting:2018tjt}. In the numerical results shown in later sections, we will follow the results of the envelope approximation and set $b = 1$. 

Matching the numerical result of the  {\it flat} space-time GW spectrum (\ref{eq:flat1p}) to Eq.~(\ref{eq:GWflat}), we have 
\bea\label{eq:flat_spec}
\frac{2 G_N}{\pi V_{\rm phy} \Delta\rho_{\rm vac}} k_p^3 \hat T^*_{ij}(k, \bk) \hat T_{ij}(k, \bk) =  \frac{\Delta \rho_{\rm vac}}{\rho_{\rm inf \star}} \left( \frac{H_\star}{\beta} \right)^2 \frac{(a+b) \tilde k^b_p k^a_p}{b \tilde k_p^{a+b} + a k_p^{a+b}} \ .
\eea
Specifically, in the IR region, 
\bea
\frac{2G_N}{\pi V_{\rm phy} \Delta\rho_{\rm vac}} \hat T^*_{ij}(0, 0) \hat T_{ij}(0, 0) =\frac{\Delta \rho_{\rm vac}}{\rho_{\rm inf \star}} \left(\frac{H_\star}{\beta}\right)^2 \times \frac{3+b}{b \tilde k_p^3} \ .
\eea
With this, the GW spectrum in the IR and the oscillatory regions can also be expressed in terms of the {\it flat} space-time GW spectrum. Thus, the GW spectrum in different regions (Eqs.~(\ref{eq:IR}), (\ref{eq:smeared}), and (\ref{eq:UV})) can be expressed in a single formula as 
\bea\label{eq:tot_spec}
\frac{d\rho_{\rm GW}}{d\log k} &=& \Delta \rho_{\rm vac} \left( \frac{1}{\Delta \rho_{\rm vac}}\frac{d\rho^{\rm flat}_{\rm GW}}{d \log k_p} \right)\left[ \cE^i_0(k) \cG^f_0(k)\right]^2 \nn & & \times \left[ 1 + {\cal S}(k_p/\beta)\cos2k(\tau_\star-\tau_0) \right]\left(\frac{a(\tau_\star)}{a(\tau)}\right)^4 \ .
\eea

In order to obtain the observed GW spectrum, we calculate the ratio of today's GW energy density to the critical density. It is convenient to consider the ratio between GW energy density to the radiation energy density since they both evolves as $a^{-4}$ after re-entering the horizon. We have
\bea
\Omega_{\rm GW} (k) \equiv \Omega_R \times \frac{d\rho_{\rm GW}(\tau_R)}{d \log k} \times \frac{1}{\rho_R(\tau_R)} \ ,
\eea
where $\rho_R$ is the energy density of the radiation and $\tau_R$ is when the reheating completes. Since at the moment the reheating completes the universe is in RD, and thus $\rho_R(\tau_R)$ and $H_r$, the Hubble parameter at $\tau_R$ is connected by the Friedman equation. Therefore, we have
\bea\label{eq:OmegaGW_today}
    \Omega_{\rm GW} (k) &=& \Omega_{R} \times \frac{\Delta \rho_{\rm vac}}{\rho_{\rm inf \star}}  \left[ \cE^i_0(k) \cG^f_0(k)\right]^2 \dfrac{H_\star^2}{H_r^2} \left(  \dfrac{a_\star}{a_r} \right)^4 \nn & & \times \left[ 1 + {\cal S}(k_p/\beta)\cos2k(\tau_\star-\tau_0) \right] \times  \frac{d \rho_{\rm GW}^{\rm flat}}{\Delta \rho_{\rm vac} d {\rm log} k_p}~,
\eea
where $\rho_{\rm inf \star}$ is the total energy density during the phase transition. We will use this formula to study the properties of the GW spectrum in the next sections.

\bigskip

\section{GW as probes of the early universe history}

\subsection{GW oscillatory pattern in different inflation models}
\label{sec:alternative}

As shown in Eqs.~(\ref{eq:smeared}) and (\ref{eq:UV}), the slopes of the oscillation profile and the UV part of the GW spectrum are sensitive to the evolution of the scale factor $a$ when the modes exit the horizon. This effect is encapsulated  by the factor $\cG^f_0(k)$ defined in Eq.~(\ref{eq:hGf}). In this sub-section, we consider this effect in detailed examples. 

\subsubsection{Quasi-de Sitter inflation}\label{desitter}
Most inflation models assume a quasi-de Sitter expansion. In the general slow roll scenario, the vacuum energy is dominated by the potential energy of the inflaton field. In this case, the Hubble parameter is almost a constant, and the universe expands exponentially. During inflation, the background is described approximately by the de Sitter spacetime, and the scale factor can be written as
\begin{align}
    a (\tau)= -\frac{1}{H \tau}~. 
\end{align}
In this case, Eq.~(\ref{eq:cgreen}) with the initial condition \eqref{init2} can be solved analytically. The solution is
\begin{align}\label{eq:GIF}
\tilde {\mathcal G}^{(1)}(\eta,\eta') = \left( \frac{1}{\eta} - \frac{1}{\eta'} \right)\cos(\eta-\eta') + \left( 1 + \frac{1}{\eta\eta'} \right) \sin(\eta-\eta') \ .
\end{align}
In the region that $(|\eta'|\gg 1)$ and $(\eta\rightarrow 0)$ we have
\begin{align}
    \frac{\tilde {\mathcal G}^{(1)}(\eta,\eta')}{a(\eta)}\bigg|_{\eta\rightarrow 0 } = \bigg( - \frac{H}{k} \bigg) \cos(\eta')~,
\end{align}
which, by comparing with \eqref{eq:hGf}, gives
\begin{align}\label{quasidSG0f}
    \tilde{\mathcal G}_0^f = \bigg( - \frac{H}{k} \bigg)~~{\rm and} \quad \tau_0 = 0~. 
\end{align}

\subsubsection{Power Law Inflation}
In the power law inflation scenario~\cite{Lucchin:1984yf}, the scale factor can be written as 
\begin{align}\label{background}
a(t) = \bar{a}\left(\frac{t}{\bar{t}} \right)^p~ = \bar{a} \bigg(\frac{\tau}{\bar{\tau}}\bigg)^{\frac{p}{1-p}}~,
\end{align}
where $\bar{a}$ and $\bar{t} (\bar{\tau})$ are some reference scale factor and reference time, which will drop out in the final result. Inflation requires $p>1$. 
The universe approaches de Sitter spacetime at $p\rightarrow \infty$. This parameterization can also be applied to alternatives to inflation. 
In this scenario, we have 
\begin{align}\label{eq:tp-H}
	H\equiv \frac{\dot a}{a} = \frac{p}{t}~ =\dfrac{p}{(1-p) \bar{a} } \dfrac{\tau^{\frac{1}{p-1}}  }{ \bar{\tau}^{\frac{p}{p-1}}  }= 	-\dfrac{p}{(p-1)  } \dfrac{1}{\bar{a}  \bar{\tau}}  \left( \dfrac{a}{\bar{a}}  \right)^{-\frac{1}{p}} ~.
\end{align}
The solution to the Green's function~\eqref{eq:cgreen} with the initial condition~\eqref{init2} is then
\begin{align} \label{greenfunctions}
\tilde {\mathcal G}^{(1)}(\eta,\eta') = \frac{\pi}{2} \sqrt{\eta \eta'}   \bigg( J_{\alpha} (-\eta)  Y_\alpha (-\eta') - J_{\alpha} (-\eta')  Y_\alpha (-\eta)  \bigg)~,
\end{align}
where $J_\alpha(x)$ and $Y_\alpha(x)$ are the Bessel functions of the first and second kind, respectively. $\alpha$ is defined as
\begin{align}
	\alpha \equiv \frac{3}{2} + \frac{1}{p-1} ~,
\end{align}
In the limit that $|\eta'| \gg 1$ and $|\eta| \rightarrow 0$ we have
\begin{align}\label{eq:}
\frac{\tilde {\mathcal G}^{(1)}(\eta,\eta')}{a(\eta)}\bigg|_{\eta\rightarrow 0 } =    \left( \dfrac{p}{p-1} \right)^{-\frac{p}{p-1}} \bar{a}^{-1}   
\left(   \dfrac{  k  }{ \bar{a} \bar{H}}  \right)^{-\frac{p}{p-1}}
\dfrac{  2^{\frac{p}{p-1}} }{\sqrt{\pi}}  \cos \bigg[\frac{\pi}{2-2p}-\eta'\bigg] \Gamma \bigg( \frac{3}{2} + \frac{1}{p-1} \bigg)~,
\end{align}
where $\bar H$ is the Hubble expansion rate at $\tau= \bar{\tau}$. Comparing with~\eqref{endofinf}, we obtain 
\begin{align}\label{eq:Gf0}
    \tilde{\mathcal G}_0^f (k) =  \left( \dfrac{p}{p-1} \right)^{-\frac{p}{p-1}} \bar{a}^{-1}   
\left(   \dfrac{  k  }{ \bar{a} \bar{H}}  \right)^{-\frac{p}{p-1}} \frac{2^{\frac{p}{p -1}}}{\sqrt{\pi}} \Gamma \bigg( \frac{3}{2} + \frac{1}{p-1} \bigg) ~, \quad k \tau_0 = \frac{\pi}{2-2p}~. 
\end{align}
Therefore, we have $\cG^f_0 \sim k^{- \frac{p}{p-1}}$. This scaling behavior can be used to distinguish the $t^p$ inflation from the quasi-de Sitter inflation.

\subsection{GW spectrum influenced by evolution after inflation}
\label{sec:later}
In this subsection, we focus on the evolution of the Universe after inflation, and its impact on the GW spectrum. From the discussion in Sec.~\ref{subsec:osc}, this effect is encapsulated by the factor $\cE^i_0(k)$, which depends solely on  the post inflationary evolution of the universe when the GW modes re-enter the horizon.

The scale factor of the the post inflationary evolution can be parameterized as 
\begin{equation}
a(t) \propto t^{\tilde p} \ .
\end{equation}
The value of $\tilde p$ can be calculated from the equation of state of the dominant component of the Universe. There are various possibilities, such as RD, MD, and cosmic string domination~\cite{Kibble:1976sj,Vilenkin:1984rt}, domain wall domination~\cite{Vilenkin:1981zs,Preskill:1991kd,Gleiser:1998na}, kination~\cite{Spokoiny:1993kt,Peebles:1998qn}, etc. We list several examples in Table~\ref{examplepost}. In the case of cosmic string domination and domain wall domination, we assume that the strings and domain walls decay before the BBN. 
\begin{table}
\begin{center}
  \begin{tabular}{ | l | c | r | c | c | }
    \hline
      & $w$ & $\rho(a)$ & $\tilde p$ & $\tilde \alpha$ \\ \hline
    kination & 1 & $a^{-6}$ & 1/3 & 0 \\ \hline
    RD & 1/3 & $a^{-4}$ & 1/2 & -1/2 \\ \hline
    MD & 0 & $a^{-3}$ & 2/3& -3/2 \\ \hline
    Cosmic string & -1/3 & $a^{-2}$ & 1 & $\infty$ \\ \hline
    Domain wall & -2/3 & $a^{-1}$ & 2 & 5/2\\ \hline
    $\Lambda$ & -1 & $a^{0}$ & $\infty$ & 3/2\\ \hline
  \end{tabular}
  \caption{\label{examplepost} Examples of post-inflationary scenarios, the corresponding equation of state $w$, the evolution of the energy density $\rho(a)$,  and the  corresponding $\tilde p$ ($a \propto t^{\tilde p} $) parameter. For later convenience, we also define $\tilde\alpha = 3/2 + 1/(\tilde{p} -1)  $. The last two rows correspond to accelerated expansion of the universe. However, in some scenarios, after inflation ends, the universe might still enter into a second stage of accelerated expansion period, e.g. the RD-$t^{\tilde p}$-RD post inflationary scenarios.}
\end{center}
\end{table}

Requiring that the scale factor $a$ and the Hubble expansion rate $H$ are continuous at the end of inflation ($a(\tau_{\rm end}) \equiv a_{\rm end }$, and $H(\tau_{\rm end}) \equiv H_{\rm end}$), the scale factor $a$ for $\tau > \tau_{\rm end}$ can be written as
\bea
a(\tau) =  f(\tau-g\tau_{\rm end})^{\frac{1}{2}-\tilde \alpha }~,
\eea
where
\bea
\tilde\alpha \equiv \dfrac{3}{2} + \dfrac{1}{\tilde{p}-1} \ ,
\eea
and 
\begin{align}
	g & = 1+ \frac{-1+2\tilde \alpha}{2 a_{\rm end} H_{\rm end} \tau_{\rm end}} ~, \quad
	f  = 2^{\frac{1}{2}- \tilde \alpha} a_{\rm end} \bigg( \frac{1-2\tilde \alpha}{a_{\rm end} H_{\rm end}}\bigg)^{-\frac{1}{2}+\tilde \alpha } ~.
\end{align}  
In particular, for  $\tau \gg \tau_{\rm end}$, we have
\beq
H(\tau) =\left(   \dfrac{a_{\rm end}}{a(\tau)}  \right)^{  \frac{1}{\tp}  } H_{\rm end} .  
\eeq 

Since the source GW considered in this work exists only during inflation, the Green's function in the region $\eta > \eta_{\rm end} \equiv k \tau_{\rm end}$ satisfies the homogeneous wave equation (\ref{eq:cgreen}). As discussed in Sec.~\ref{subsec:osc}, the evolution of the GW in the regime can be described by the function ${\cal E}(\eta)$ satisfying the initial condition (\ref{eq:Einitial}). The general form of ${\cal E}$ can be written as  
\begin{align} \label{greenfunctions}
\mathcal E (\eta) =  (\eta - g\eta_{\rm end} )^{\tilde \alpha} \bigg[  L_1  J_{\tilde \alpha}(\eta- g\eta_{\rm end} )+L_2  Y_{\tilde \alpha}(\eta- g \eta_{\rm end} ) \bigg]~.
\end{align} 
Since the GW modes under consideration are all far outside the horizon at the end of the inflation, we have $|\eta_{\rm end}| = k |\tau_{\rm end}| \ll1$. 
Requiring $\mathcal E\rightarrow 1$ and $\mathcal E'\rightarrow 0$ as  $\eta\rightarrow\eta_{\rm end}\ll 1$, we can fix the constants in Eq.~\eqref{greenfunctions} as 
\begin{align}\label{coefficient1}
    & L_1 = 2^{-2 +\tilde{\alpha}} \pi  \left( \dfrac{1- 2 \tilde{\alpha}}{a_{\rm end} H_{\rm end}} \right)^{1-\tilde{\alpha}  }  Y_{\tilde {\alpha} -1} \left( \dfrac{k(1- 2 \tilde \alpha)}{2 a_{\rm end} H_{\rm end}}  \right) \stackrel{|\eta_{\rm end}| \to 0}{\xrightarrow{\hspace{1cm}}}  \dfrac{1}{2} \cos (\pi  \tilde \alpha ) \Gamma (1-\tilde \alpha ) (1- 2 \tilde {\alpha})^{1-\tilde {\alpha}} \quad \\ 
    \label{coefficient2}
   & L_2 = -2^{-2 +\tilde{\alpha}} \pi  \left( \dfrac{1- 2 \tilde{\alpha}}{a_{\rm end} H_{\rm end}} \right)^{1-\tilde{\alpha}  }  J_{\tilde {\alpha} -1} \left( \dfrac{k(1- 2 \tilde \alpha)}{2 a_{\rm end}  H_{\rm end} }  \right)  \stackrel{|\eta_{\rm end} | \to 0}{\xrightarrow{\hspace{1cm}}}  -\frac{\pi  }{2 \Gamma (\tilde \alpha )}   (1- 2 \tilde {\alpha})^{1-\tilde {\alpha}}   ~.
\end{align} 
When the modes re-enter  the horizon, $\eta\gg 1$, we have
\begin{align}\label{eq:Ei0}
    \mathcal E ( \eta) =  \Gamma (\tilde\alpha )^{-1} \sqrt{\pi } 2^{\frac{1}{2}-\tilde \alpha } \csc (\pi \tilde \alpha ) \left(\eta - g\eta _{\rm end}  \right){}^{\tilde \alpha -\frac{1}{2}}
   \sin \left(\frac{\pi \tilde \alpha }{2}+\eta - g\eta _{\rm end} +\frac{\pi }{4}\right)  ,
\end{align}
from which we can read off the evolution factor $\cE^i_0$ defined in Eq.~(\ref{eq:cEi}) as
\begin{align}\label{eq:cE0i}
    \tilde {\mathcal E}_0^i (k)=\ 2^{\frac{1}{2}-\tilde \alpha } \sqrt{\pi}  f \csc (\pi  \tilde \alpha ) k^{\tilde\alpha -\frac{1}{2}}  \Gamma (\tilde \alpha)^{-1}~.
\end{align}

\subsubsection{RD}\label{RD}

 The most commonly considered post inflationary scenario is RD. In this case, $\tilde p = 1/2$, then $\mathcal E(\eta)$ evaluates to \begin{align}
    \mathcal E(\eta) = \frac{\sin \eta }{\eta }~,
\end{align}
from which we can extract 
\begin{align}\label{eq:3.21}
    \tilde {\mathcal E}_0^i = \frac{f}{k}~,\quad f= a_{\rm end}^2 H_{\rm end}~.
\end{align} 

\subsubsection{MD}\label{MD}
In the case of MD, the function $\mathcal E(\eta)$ behaves as
\begin{align}
    \mathcal E(\eta) = -\frac{3}{\eta^2} \bigg( \cos\eta - \frac{\sin\eta}{\eta} \bigg),
\end{align}
and 
\begin{align}
    \tilde {\mathcal E}_0^i = \frac{3 f}{k^2}~, \quad f= \frac{a_{\rm end}^3 H_{\rm end}^2}{4}~.
\end{align}

\subsection{Comparing different scenarios}
\label{sec:compare}

\begin{table}[t]
\centering 
	\begin{tabular}{|c|c|c|c|c|}
		\hline
		\multirow{3}{*}{UV} & & RD                         & MD & $t^{\tilde{p}}$      \\ \cline{2-5}
		 & dS                     & $k^{-b-4}$  &   $k^{-b-6}$   &      $k^{-b-\frac{2}{1 - \tilde{p} }  }    $        \\ \cline{2-5}
		 & $t^p$  &   $k^{-4-b-\frac{2}{ p  -  1}} $    & $k^{-6-b-\frac{2}{ p  -  1}} $   & $k^{-b-2\left(  \frac{1}{p-1}  + \frac{1}{1-\tilde{p}}  \right)} $   \\ \hline
		 	\hline
		 \multirow{3}{*}{Intermediate} & & RD                         & MD & $t^{\tilde{p}}$      \\ \cline{2-5}
		& dS                     & $k^{-1}$  &   $k^{-3}$   &      $k^{3 -\frac{2}{1-\tilde{p}}  }    $        \\  \cline{2-5}
		& $t^p$  &   $k^{-1 - \frac{2}{ p  -  1}} $    & $k^{-3 - \frac{2}{ p  -  1}}$   & $k^{3-2\left(  \frac{1}{p-1}  + \frac{1}{1-\tilde{p} }  \right)} $   \\ \hline
				\hline
		\multirow{3}{*}{IR}		
		& & RD                         & MD & $t^{\tilde{p}}$      \\ \cline{2-5}
		& dS                     & $k^{3}$  &   $k^{1}$   &      $k^{7-\frac{2}{1-\tilde{p}}  }    $        \\  \cline{2-5}
		& $t^p$  &   $k^{3} $    & $k^{1} $   & $k^{7-\frac{2}{1-\tilde{p}}  }     $   \\ \hline
	\end{tabular}
\caption{The slope of the GW spectrum in different frequency regimes.  In the UV regime, the GW scales as $\left[\cE^i_0(k) \cG^f_0(k)\right]^2 k^{-b} $. We use $b=1$ for the numerical results presented here. The factor $\left[\cE^i_0(k) \cG^f_0(k)\right]^2$  encodes the  effect from the propagation of the GW,  $k^{-b}$ is contributed by the UV part of the source. In the intermediate regime, it scales as $   \left[\cE^i_0(k) \cG^f_0(k)\right]^2 k^3 $, where $k^3$ is contributed by the IR part of the source. In the IR regime,  it scales as  $ \left[\cE^i_0(k)\right]^2 k^5$, where $\left[\cE^i_0(k)\right]^2 k^2$ encodes the IR effect from the propagation of the GW. The additional $k^3$ is contributed by the IR part of the source. \label{tab:three_regimes}}
\end{table}

As discussed in Sec.~\ref{sec:generic}, in the IR part of the spectrum, neither  $\cG^f_0$ nor $\hat T$  depend on $k$. Therefore, slope of the spectrum is determined completely by $k^3 |\cE^i_0|^2$. In the oscillatory part, the GW spectrum (\ref{eq:OmegaGW_today}) is determined by the factor $\left[ \cE^i_0(k) \cG^f_0(k)\right]^2 $. 
Inserting the expressions of $\cG^f_0$ and $\cE^i_0$ from Eqs.~(\ref{eq:Gf0}) and (\ref{eq:Ei0}), we can factor out the $k$ dependence of the profile of the oscillatory part of the spectrum. Using $a = t^p$ and $a = t^{\tilde p}$ to parameterize the inflationary and post-inflationary evolutions of the universe,  we have
\bea
k^3 \left[\cE^i_0(k) \cG^f_0(k)\right]^2 \sim k^{3 - \frac{2p}{p-1} + \frac{2\tilde p}{\tilde p - 1}} \ . 
\eea
The UV part of the spectrum also depends on the details of the phase transition. As discussed in Sec.~\ref{sec:intro}, we focus on scenarios in which the energy density of the plasma is subdominant compared to the vacuum energy. As a result, the GWs are mainly induced by the collision of bubble walls. Numerical simulations show that in the UV region, the GW in flat space-time behaves as $k^{-b}$. 
The slope in the UV region of the GW spectrum can be written as
\bea
k^{-1} \left[\cE^i_0(k) \cG^f_0(k)\right]^2 \sim k^{-b - \frac{2}{p-1} - \frac{2}{1 - \tilde p }} \ .
\eea
The slopes of the UV, intermediate and IR regions of the GW spectrum for typical evolution models are listed in Table~\ref{tab:three_regimes}.

\subsubsection{$t^p$ vs quasi-de Sitter Inflation}

\begin{figure}[h!]
\centering
\includegraphics[height=2.2in]{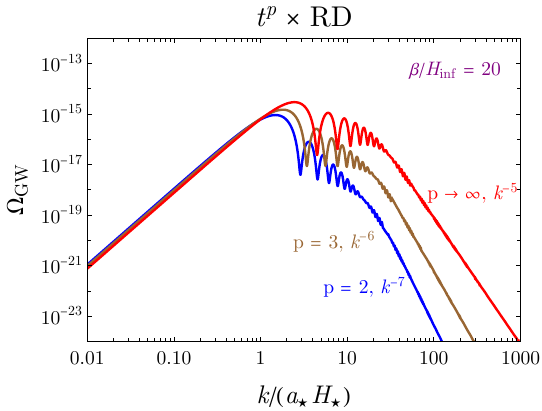}\quad 
\includegraphics[height=2in]{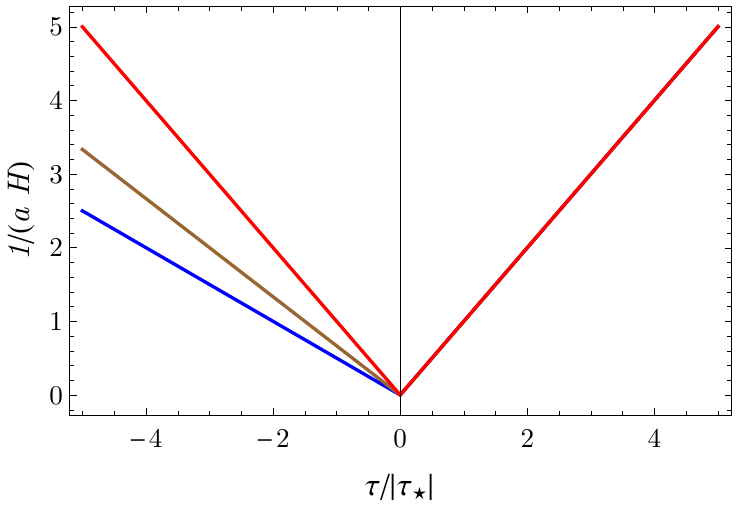}
\caption{The left panel shows the comparison of GW signal from quasi-de Sitter inflation (red) and $t^p$ inflation with $p=2$ (blue). The right panel shows the corresponding co-moving horizon $1/(a H)$ as a function of conformal time $\tau$ in these scenarios (with the same color legend).  The three curves do not coincide in the region $\tau>0$ on the right panel but their difference is almost invisible from the plot. Both the UV and the oscillatory part of the GW spectrum in the three different scenarios are distinct. }\label{fig:tp_dS}
\end{figure}

Substituting the $\cE^i_0$ in $\cG^f_0$ into the the GW spectrum (\ref{eq:OmegaGW_today}), we obtain the spectrum for general choices of $p$ and $\tilde p$. For comparison of the effect from different values of $p$, we fix the evolution right after the inflation to be RD ($\tilde p = 1/2$).
 The GW spectra for different values of $p$ are shown in Fig.~\ref{fig:tp_dS}, where the brown and red curves are for $p = 2$ and $3$. Both the slopes of oscillatory part and the UV part are changed compared to the quasi-de Sitter case (shown as the red curve with $p\rightarrow\infty$).

\subsubsection{de Sitter-$t^p$-de Sitter transition during inflation}

We consider here a scenario in which the inflationary evolution involves three stages. We assume the first stage ($\tau < \tau_{{\rm dS}_1}$) is the usual quasi-de Sitter inflation, with $H= H_{{\rm dS}_1}$. In the second stage, the vacuum energy decreases and the evolution deviates from the quasi-de Sitter. We can parameterize the evolution in this stage with $a \propto t^p$. 
The third stage, also quasi-de Sitter,  starts 
at $\tau = \tau_{{\rm dS}_2}$. In this stage, the universe experiences another quasi-de Sitter inflation phase with $H= H_{{\rm dS}_2}$. This transition can deform the GW spectrum produced by a first-order phase transition happened during the first quasi-de Sitter inflation period\footnote{The transition can in principle leave imprints on the CMB if it happened during the epoch when the corresponding modes were exiting the horizon. Otherwise, the GW signal discussed here would be the only window into such a transition. }. 
The scale factor of each periods can be written as
\begin{align}\label{eq:ataudStpdS}
	a(\tau) = \begin{cases}
	-\left( H_{{\rm dS}_1}(\tau+\frac{\tau_{{\rm dS}_2}-\tau_{{\rm dS}_1}}{p}) \right)^{-1},\quad H=H_{{\rm dS}_1},\quad \tau<\tau_{{\rm dS}_1} \\      \  \\
	  a(\tau_{{\rm dS}_1}) \left( 1+ \dfrac{\tau-\tau_{{\rm dS}_1}}{ \tau_{{\rm dS}_1}+ \tau_{{\rm dS}_2}/(p-1)  }  \right)^{\frac{p}{1-p}}, \quad
	H = H_p(\tau) ,
	\quad  \tau_{{\rm dS}_1}  <\tau < \tau_{{\rm dS}_2}
	\\ \ \\
	-(H_{{\rm dS}_2} \tau)^{-1}, \quad   H = H_{{\rm dS}_2},  \quad
	\tau_{{\rm dS}_2} <\tau < 0	
	\end{cases},
\end{align} 
where
\begin{align}
	H_p(\tau)  = H_{{\rm dS}_1} \left( \dfrac{\tau + (\tau_{{\rm dS}_2})/(p-1) }{ \tau_{{\rm dS}_1} + (\tau_{{\rm dS}_2})/(p-1) }\right)^{\frac{1}{p-1}}~, \\
	H_{{\rm dS}_2}=  H_{{\rm dS}_1} \left( \dfrac{p \, \tau_{{\rm dS}_2}}{(p-1)\tau_{{\rm dS}_1}  + \tau_{{\rm dS}_2} } \right)^{\frac{1}{p-1}} \ .
\end{align} 
The mode crosses the horizon when its physical momentum $k/a(\tau)$ is comparable to the Hubble parameter. From the above equations, we can obtain the ratio $k/(a H)$ at different periods of the evolution of the universe that
\begin{align}\label{eq:dStpdS-kaH}
	\dfrac{k}{a(\tau) H(\tau)} = \begin{cases}
	-k \left(\tau+\dfrac{\tau_{{\rm dS}_2}-\tau_{{\rm dS}_1}}{p}  \right),
	\quad \tau<\tau_{{\rm dS}_1}  \vspace{0.2cm}  \\    
	- k \dfrac{(p-1)\tau + \tau_{{\rm dS}_2}}{p}, \quad
	H = H_p(\tau) ,
	\quad  \tau_{{\rm dS}_1} <\tau < \tau_{{\rm dS}_2} \vspace{0.2cm}
	\\ -k \tau, \quad
	\tau_{{\rm dS}_2} <\tau < 0	
	\end{cases}
\end{align} 
\begin{figure}[h!]
\centering
\includegraphics[height=1.9in]{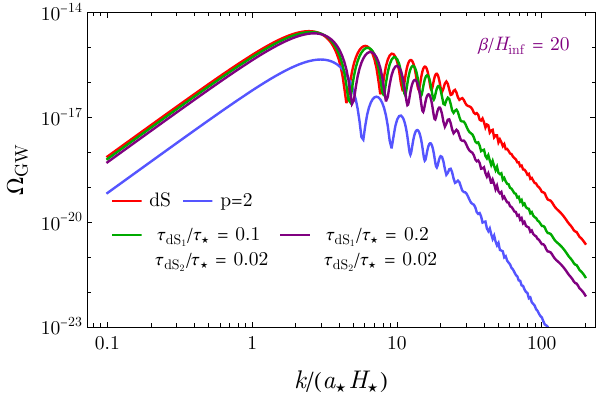}
\includegraphics[height=1.9in]{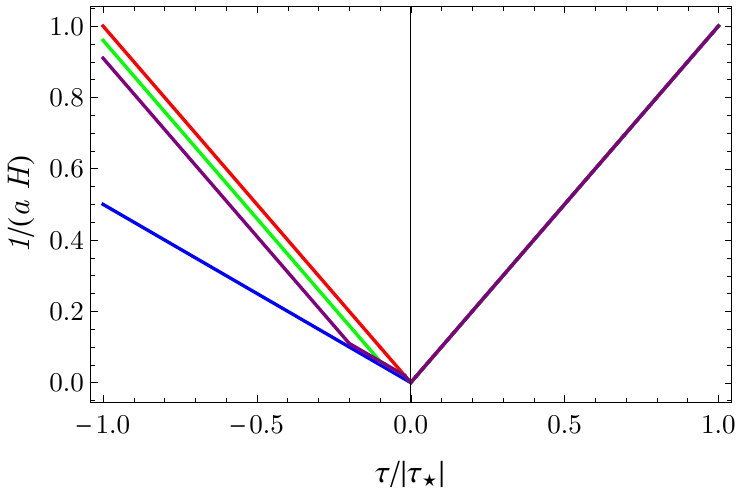}
\caption{The left panel shows observed GW spectrum for different $k$ modes in several different inflationary scenarios. The right panel shows the evolution of the co-moving horizon as a function of conformal time. Different colors denotes different cosmic evolution histories before reheating, including quasi-de Sitter (red), $t^2$ inflation (blue), and dS-$t^p$-dS scenarios with different parameters (green and purple lines). }
\label{fig:dStpdS1}
\end{figure}
Fig.~\ref{fig:dStpdS1} shows the observed spectra of different GW modes as functions of $k/(a_\star H_\star)$. 
The mode with $k/(a_\star H_\star) = 1$ crosses horizon during the phase transition. 
The Green curve describes the case $\tau_{{{\rm dS}_1}}/\tau_\star  = 0.1$ and $\tau_{{{\rm dS}_2}}/\tau_\star = 0.02$ with $p=2$. In this case, according to Eq.~(\ref{eq:dStpdS-kaH}), modes with $1<k/(a_\star H_\star) < 17$ exit horizon in the first quasi-de Sitter stage, whereas modes with $17 < k/(a_\star H_\star) < 50$ exit the horizon in the intermediate $t^p$ stage. Modes with shorter wavelength, $k/(a_\star H_\star) > 50$, exit the horizon in the second quasi-de Sitter stage. 
For comparison, in Fig.~\ref{fig:dStpdS1}, we also show the GW spectra in the quasi-de Sitter inflation and $t^p$ inflation with $p=2$ with the red and blue curves, respectively. One can see that the slopes of profiles of the green curve is in parallel to the red curve in the regions of $k/(a_\star H_\star) < 16.7$ and $k/(a_\star H_\star)  > 50$. Whereas in the region $16.7 < k/(a_\star H_\star) < 50$, the green curve is in parallel to the blue curve. The purple curve shows that case that $\tau_{{\rm dS}_1}/\tau_\star  = 0.2$ and $\tau_{{\rm dS}_2}/\tau_\star  = 0.02$. In this case the transitions happen at $k/(a_\star H_\star) \approx 9$ and 50.

\subsubsection{$t^{\tilde{p}}$-RD transition in post-inflationary evolution}
\label{sec:tpRD}

After inflation, the universe may go through some intermediate stages before finally entering the RD regime. This may significantly change the GW spectrum. If the reheating  lasts less than one e-fold, it is reasonable to assume that most of the energy in the inflaton field goes into radiation.  We call this scenario immediate reheating. Instead of this simple picture, the universe could experience an intermediate stage before entering RD. 
The nature of such an intermediate period is model-independent. For example, the kinetic energy of the inflaton field may dominate the universe right after the inflation. This leads to the $t^{1/3}$ expansion, which is known as the kination domination (KD). Another example is that the inflaton field may oscillate for a while after the end of inflation. Its equation of state would be  similar to the pressure-less dust and will lead to a MD era. Here, we use a general $t^{\tilde{p}}$-RD scenario to parametrize such an intermediate transition stage after inflation. We denote $\tau_{R}$ to be the conformal time at the $t^{\tilde{p}}$-RD transition. Then the case $\tau_{R} = \tau_{\rm end}$ refers to the immediate reheating scenario. In the case that the $t^\tp$ stage lasts significantly longer than one e-fold, we have $\tau_{R}\gg \tau_{\rm end}$, and therefore, from Eq.~(\ref{eq:app_tp_RD_aH}) 
\begin{equation}\label{eq:tpRDa}
    \dfrac{a_{R}}{a_{\rm end}} = \brr{\dfrac{\aend \He \tau_{R}}{\tilde{\omega}}}^{\tilde{\omega}} \ , \;\; \quad \dfrac{H_{R}}{H_{\rm end}} = \brr{\dfrac{\aend \He \tau_{R}}{\tilde{\omega}}}^{-\tilde{\omega}-1} \ , \;\; \quad
    \dfrac{k}{a_{R} H_{R}} = \tomega\, k\, \tau_{R}
\end{equation}
where $\tomega = \tp/(1-\tp), a_{R}=a(\tau_{R})$ and $H_{R} = H(\tau_{R})$.  From the last relation in Eq.~(\ref{eq:tpRDa}), we know that modes with $k>  \brr{\tilde{\omega} \tau_{R}}^{-1}$ re-enter the horizon during the intermediate $t^\tp$ stage, whereas modes with $k < \brr{\tilde{\omega} \tau_{R}}^{-1}$ re-enter the horizon at the RD stage. The detailed calculations of the Green's function ${\cal E}$ and the factor ${\cE^i_0}$ are presented in the appendix~\ref{app:tp-RD}. 
\begin{figure}[h!]
\centering
\includegraphics[height=2.1in]{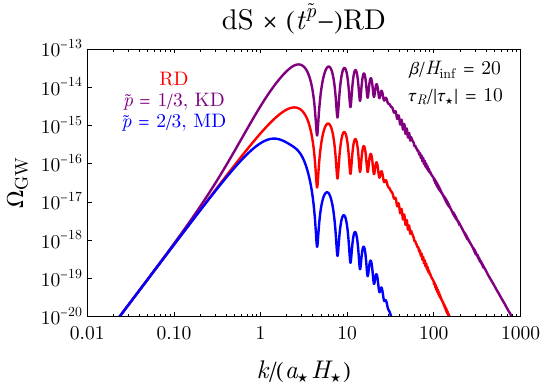}
\includegraphics[height=1.9in]{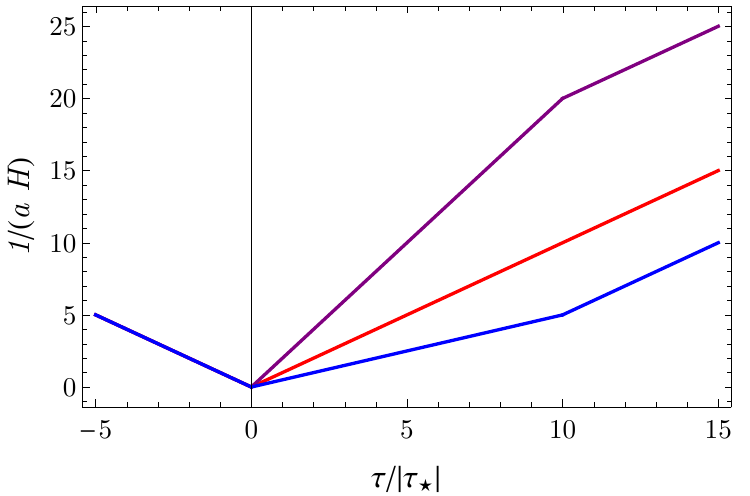}
\caption{Left: GW spectrum from first-order phase transition during the quasi-de Sitter inflation, in scenarios with a $t^{\tp}$ intermediate stage before the RD era and the inflation. The blue and purple curves are for MD and KD intermediate stages, respectively. As a comparison, the GW spectrum for the instantaneous reheating scenario is also shown as the red curve. For all the three scenarios $\tau_{R}$ is fixed to be $10 |\tau_\star|$. Right: the evolution of $1/(aH)$ for the scenarios shown in the left panel. 
}\label{fig:dS_tp_RD_with_p}
\end{figure}

\begin{figure}[h!]
\centering
\includegraphics[height=2.1in]{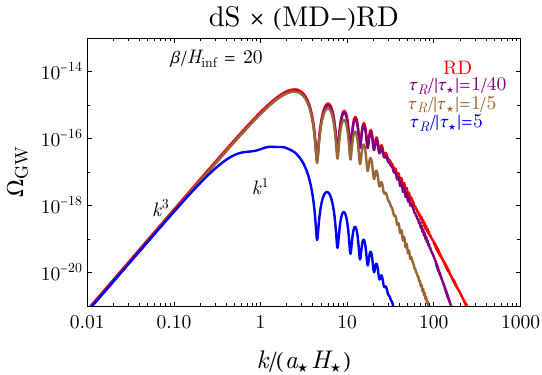}
\includegraphics[height=1.9in]{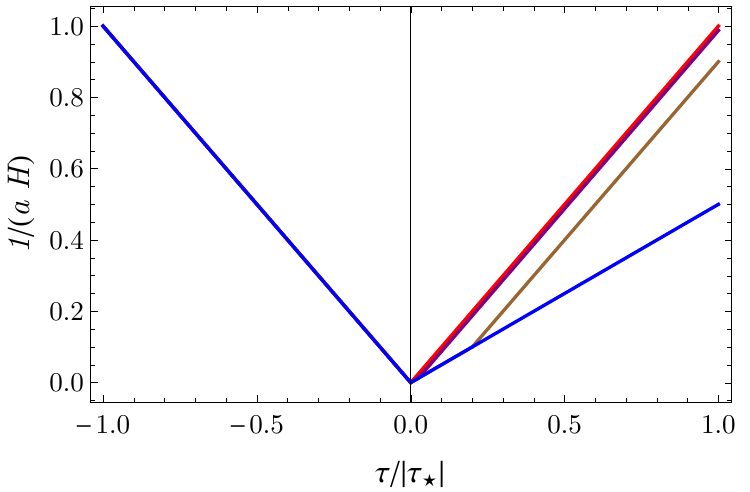}\\ \vspace{0.2cm}
\includegraphics[height=2.1in]{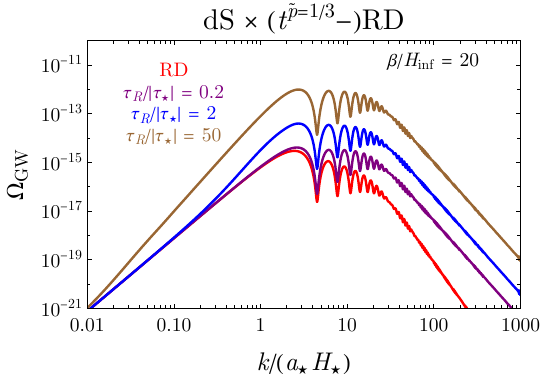}
\includegraphics[height=1.9in]{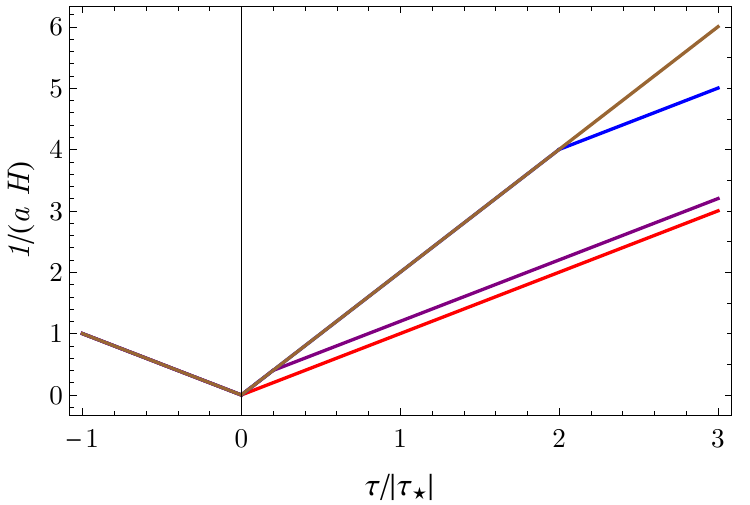}
\caption{Left: GW spectrum from first-order phase transition during the quasi-de Sitter inflation, in scenarios with an MD intermediate stage before the RD era in the left upper panel and with an KD intermediate stage before the RD era in the left lower panel. The brown, blue and purple curves are for different $\tau_R/|\tau_\star|$ settings. As a comparison, the GW spectrum for the instantaneous reheating scenario is shown as the red curve. For all the  scenarios $\beta/H_{\rm  inf}$ is fixed to be 20. Right: the evolution of $1/(aH)$ for the scenarios shown in the left panel. }\label{fig:MDRD}
\end{figure}
The purple and blue curves in the left panel of Fig.~\ref{fig:dS_tp_RD_with_p} show the GW spectra in the scenarios with KD and MD stages between inflation and the RD stage, respectively. As a comparison, the scenario with instantaneous reheating is also shown by the red curve. In the plot the value of $\beta/H$ is fixed to be 20, and the value of $|\tau_R/\tau_\star|$ is fixed to be 10. This value is chosen such that the whole oscillatory part of the GW spectrum re-enters the horizon during the $t^\tp$ stage. Indeed, the value of $k/(a_\star H_\star)$ is just $(\tau_\star/\tau_{R})\tilde\omega$, which equals 0.2 and 0.05 for the MD and KD intermediate scenarios. These are the regions where the blue and purple curves start to deviate from the red curve. 
In Fig.~\ref{fig:dS_tp_RD_with_p}, one can also see that the deep IR region of the spectra with different choices of $\tp$ are all coincident with each other. This is due to that all the deep IR modes are out-of-horizon upon their production. Therefore, their amplitudes does not red-shift until they re-enter the horizon. The detailed reason will be explained in Sec.~\ref{sec:strength}. 
The slopes for the MD and KD intermediate scenarios in each region can be read off from Table~\ref{tab:three_regimes} and are shown explicitly in Fig.~\ref{fig:dS_tp_RD_with_p}. The slope in the IR region is solely determined by the post-inflationary evolution when the modes re-enter the horizon, which are $k^3$, $k^1$ and $k^4$ for the RD, MD and KD cases, respectively. Therefore, we can get that, relatively, if there is a KD intermediate stage after inflation, the GW signal will be enhanced compared to the instantaneous reheating scenario. Whereas if there is intermediate MD stage, the GW signal will be suppressed. It is worth mentioning that the profile of the oscillatory region in scenario with the KD intermediate stage is flat for the parameters we choose, as we can see from the purple curve in Fig.~\ref{fig:dS_tp_RD_with_p}. This means if  a GW detector cannot resolve the oscillatory pattern, it will see a flat GW spectrum.  
The corresponding plot of the evolution of the factor $1/(a H)$ for the each scenario is shown in the right panel. 

In Fig.~\ref{fig:MDRD}, we show the GW spectrum in the scenarios with MD and KD intermediate stage with different values of $\tau_R / |\tau_\star|$. As a comparison, the instantaneous reheating scenario is also shown by the red curves. One can see that strength of the signal is determined by the duration of the intermediate stage. In the KD intermediate scenario, the longer the KD stage is, the larger the peak GW signal is. On the contrary, in the MD intermediate scenario, the longer the MD stage is, the smaller the signal is.

\subsubsection{RD-$t^{\tp}$-RD transition in post-inflationary evolution}

The post-inflation evolution of the universe can be even more complicated. For example, during the RD era, if the particle content contains a long lived particle (with lifetime much longer than Hubble scale at the temperature roughly equal to the mass of this particle), the universe will undergo a temporary MD era between two RD eras.
During the post-inflation expansion, the universe may also undergo a second order phase transition. As a consequence, the universe may be dominated temporarily by topological defects, such as cosmic strings or domain walls. 

All these cases can be  described by a RD-$t^{\tilde p}$-RD scenario, in which the post-inflation epoch includes a RD era, followed by an intermediate stage with $a \propto t^{\tilde p}$, and then back to a second RD era. Different $\tilde{p}$ corresponds to different evolution models. For example, $\tilde p = 2/3$ and $\tilde p= 1$ describe the evolution dominated by a species of  long lived particles and long lived cosmic strings, respectively. The values of $\tilde p$ for different models are shown in Table~\ref{examplepost}. $\tilde{p} > 1$ corresponds to the inflationary case with the shrinking comoving Hubble sphere. In the critical case $\tilde{p} = 1$, the physical momentum red-shift in the rate as the Hubble expansion rate. As a result, there are no modes crossing horizon in the $\tp = 1$ stage. $\tilde{p}<1$ leads to an expanding comoving sphere and more and more modes will re-enter the horizon. 

\begin{figure}[th!]
\centering
\includegraphics[height=2.1in]{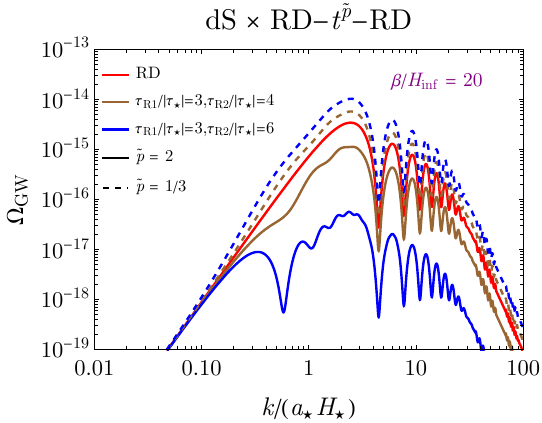}
\includegraphics[height=1.9in]{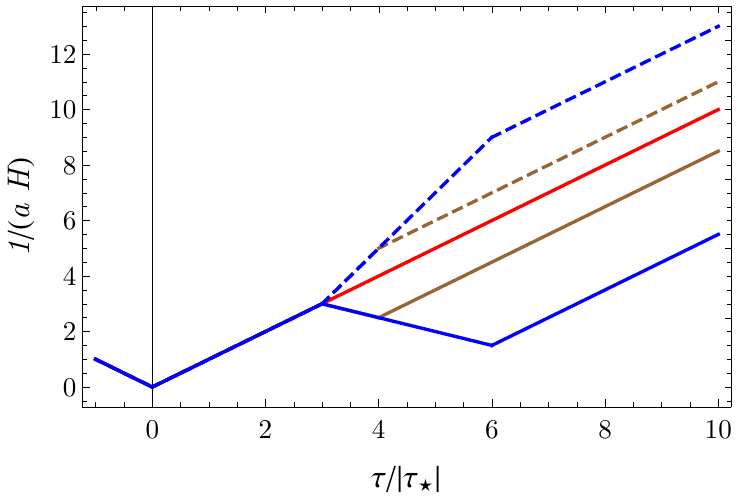}
\caption{GW spectrum from first-order phase transition during the quasi-de Sitter inflation, with the post inflationary RD$-t^{\tp}-$RD scenario. Possible intermediate stages with $\tilde{p} = 2$ (solid) $\tilde{p} = 1/3$ (dashed) which corresponding to the domain wall and kination cases in Table~\ref{examplepost} are included.  The red curve corresponds to the dS$\times$RD scenario. The brown and blue curves are for different ratios of $\tau_{{R}_1}/|\tau_\star|$ and $\tau_{{R}_2}/|\tau_\star|$, respectively. The solid line is for $\tp=2$ while the dashed line is for $\tp=1/3$. The right panel shows the comoving horizon evolution.}\label{fig:RDwRDplot1}
\end{figure}

\begin{figure}[th!]
\centering
\includegraphics[height=2.1in]{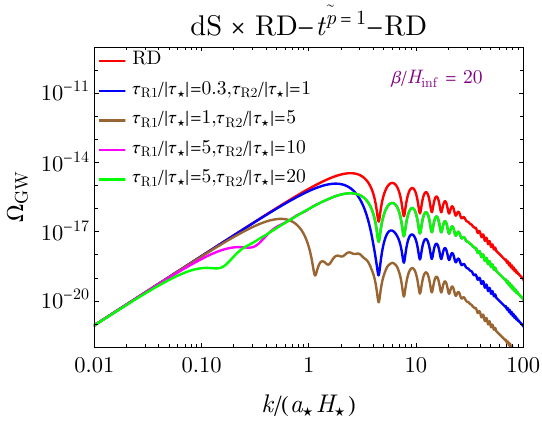}
\includegraphics[height=1.9in]{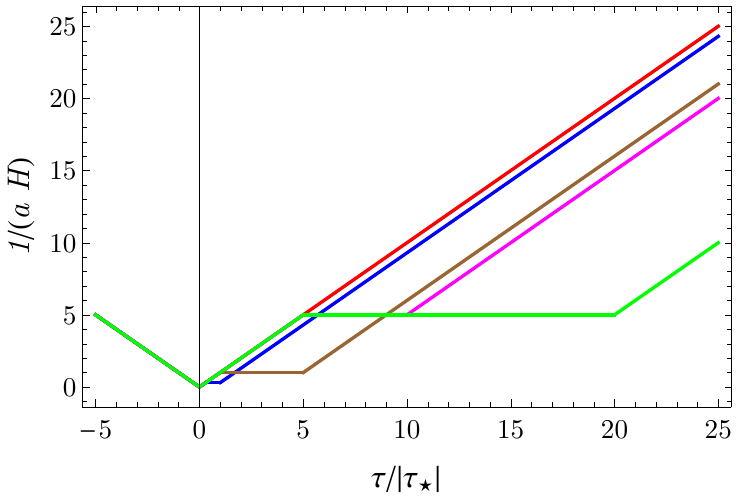}
\caption{GW spectrum from first-order phase transition during the quasi-de Sitter inflation, with the post-inflationary RD$-t^{\tilde{p}=1}-$RD scenario. The intermediate $t^{\tilde{p}=1}$ stage corresponds to the cosmic string dominated era. The red curve corresponds to the dS$\times$RD scenario. The curves with other colors are for different ratios of $\tau_{{R}_1}/|\tau_\star|$ and $\tau_{{R}_2}/|\tau_\star|$, respectively.}\label{fig:RDtp1RDplot}
\end{figure}

On the left panel of  Fig.~\ref{fig:RDwRDplot1}, we plot the GW signal with an intermediate stage dominated by domain wall ($\tilde{p} = 2$, solid line) and kination ($\tilde{p} = 1/3$, dashed line). $\tau_{{R}_1}$ and $\tau_{{R}_2}$ denote the starting and ending conformal time of the intermediate $t^\tp$ stage. The brown and blue curves correspond to different values of $\tau_{{R}_2}/|\tau_\star|$. For both cases, we choose $\tau_{R_1}/|\tau_\star| = 3$. Then, we choose $\tau_{R_2}/|\tau_\star| = 4$ and 6 for the brown and blue curves, respectively. As discussed in Sec.~\ref{sec:tpRD}, with the choice $\tau_{R_1}/|\tau_\star| = 3$, modes in the oscillatory and UV parts of the spectrum re-enter the horizon before the $t^{\tp}$ stage. Therefore, just as we see in Fig.~\ref{fig:RDwRDplot1}, the oscillatory and UV parts of the spectra are all in parallel with each other. The intermediate evolution of the universe leaves imprints only in the IR part of the spectrum. For example, as shown by the dashed blue and brown curves the slope in the IR part is 4 in the KD intermediate case as discussed in Sec.~\ref{sec:tpRD}. In the case of domain wall domination, the expansion of the is accelerating, and the universe is indeed inflating. Therefore, as discussed in Sec.~\ref{sec:2point1}, an additional oscillatory pattern will show up in the spectrum, which are explicitly shown by the wiggles around the $k/(a_\star H_\star)$ region in the brown and blue solid curves. The red panel of Fig.~\ref{fig:RDwRDplot1} shows the evolution of the size of the horizon of each scenarios.

First, for the modes which re-enter the horizon before $\tau_{R_2}$, the signal is enhanced (suppressed) for ${\tilde p} = 1/3$ (${\tilde p} = 2$) in comparison with RD.  There is also a suppression for the UV and part of the intermediate modes. This is because they have already entered the horizon and are subject to the $a^{-4}$ suppression. For the case in which modes with $k\sim 1/|\tau_\star|$ re-enter horizon at the matter dominated epoch, the suppression sets in for a lower frequency range, which is also expected from the same argument. On the left panel of Fig.~\ref{fig:RDwRDplot1}, we also show the GW signal with $k \sim 1/|\tau_\star|$ re-enters horizon at the first radiation dominated epoch (blue). In this case, the suppression begins in the IR regime. 
$\tilde{p} =1$ case is unique cause $\tilde{\omega} = \infty$. We discuss the evolution in this case in Sec.~\ref{sec:RDtp1RD} and show the GW signal in Fig.~\ref{fig:RDtp1RDplot}. 

\section{Observed GW Signals}
\label{sec:signal}

In this section we focus on the properties of the observed signals at GW observatories. 
The  GW spectrum is shown in Eq~(\ref{eq:OmegaGW_today}). The overall strength of the signal is proportional to $\Delta \rho_{\rm vac}/\rho_{\rm inf \star}$. In the numerical results presented here, we made a somewhat conservative choice of $\Delta \rho_{\rm vac}/\rho_{\rm inf \star} =0.1$. The vacuum energy is chosen to be small in order to avoid influence to the dS background. In principle, this fraction can certainly be larger. 

\subsection{Strength of the peak signal}
\label{sec:strength}

The peak of the GW signal and its corresponding frequency depend on the spectral shape. From Eq.~(\ref{eq:flat_spec}) and Eq.~(\ref{eq:tot_spec}),the signal strength is proportional to
\bea\label{eq:ratiop}
\left( \frac{H_\star}{\beta} \right)^2 \frac{(a+b) \tilde k^b_p k^a_p}{b \tilde k_p^{a+b} + a k_p^{a+b}} \left[ \cE^i_0(k) \cG^f_0(k)\right]^2  \left(H_\star^2/H_r^2\right) \left(  a_\star/a_r \right)^4 \ . 
\eea
The dependence on the wave number $k$ of each part of the GW spectrum for different scenarios are shown in Table~\ref{tab:three_regimes}. In general, we can parameterize the scale factor $a$ during inflation as $t^p$ with $p > 1$, and after inflation as $t^{\tilde p}$. The spectrum in the UV region can be parameterized as $k^{i_{\rm UV}}$, with $i_{\rm UV} \equiv - b - 2[(p-1)^{-1} + (1 - \tilde p)^{-1}]$, which is always negative. In the intermediate region, the spectrum is $k^{i_{\rm OSC}}$, with  $i_{\rm OSC} \equiv 3 - 2[(p-1)^{-1} + (1-\tilde p)^{-1}]$. $i_{\rm OSC}$ can  be either positive or negative. In the case of quasi-de Sitter inflation, we have $p \rightarrow \infty$. Therefore, $i_{\rm OSC} < 0$ if $\tilde p > 1/3$. For example, this would be the case for both MD and RD, as shown in Table~\ref{examplepost}. At the same time,  we have $i_{\rm OSC} = 0$ for kination domination. For the IR part of the spectrum, parameterized as $k^{i_{\rm IR}}$, we have  $i_{\rm IR} = 7 - 2(1 - \tilde p)^{-1}$. Therefore, $i_{\rm IR} > 0$ if $\tilde p < 5/7$. As we can see from Table~\ref{examplepost}, this condition is always satisfied in the standard cases discussed in the literature. Thus,  the position of the global maximum of the observed GW spectrum is determined by $i_{\rm OSC}$. More specifically,
\begin{enumerate}
\item[${\bullet }$]
If $i_{\rm OSC} < 0$, the global maximum is at the transition between the IR part and the oscillatory part, where we have $k_p \approx H_\star$. This is the case shown in Figs.~\ref{fig:tp_dS},\ref{fig:dStpdS1},\ref{fig:MDRD} and \ref{fig:RDwRDplot1}. Hence, the height of the global maximum can be estimated by substituting $k_p = H_\star$ to Eq.~(\ref{eq:flat_spec}). Then we have
\bea
\frac{d\rho^{\rm flat}}{\Delta\rho_{\rm vac} d\log k_p} \sim \left(\frac{H_\star}{\beta}\right)^5 \ .
\eea
Therefore, in this case, the GW signal strength at the global maximal can be estimated as
\bea\label{eq:OmegaM}
\Omega_{\rm GW}^{\rm max} \sim \Omega_R \times \brr{\frac{\Delta\rho_{\rm vac}}{\rho_{\rm inf \star}} }^2 \times \left(\frac{H_\star}{\beta}\right)^5 \tilde \Delta \times{ F} (H_\star/H_r, a_\star/a_r, \cdots) \ ,
\eea
where the dimensionless function ${F}$ depends on the details of the evolution of the Universe from the phase transition to reheating. For the simplest case, quasi-de Sitter inflation followed by instantaneous reheating, detailed calculation shows that ${F} = 1$. Therefore, one can use the value 
\bea
\Omega_R \times \brr{\frac{\Delta\rho_{\rm vac}}{\rho_{\rm inf \star}}}^2 \times \tilde\Delta\times \left(\frac{H_\star}{\beta}\right)^5 \approx 10^{-13} \times \brr{\frac{\Delta\rho_{\rm vac}/\rho_{\rm inf \star}}{0.1}}^2  \times \left( \frac{H_\star/\beta}{0.1} \right)^5 \ . 
\eea
as a benchmark value for the strength of the GW signal. 
\item[${\bullet }$]
If $i_{\rm OSC} > 0$, the global maximum is at the transition between the oscillatory part and the UV part with $k_p \sim \beta$. To estimate the strength of the signal in this transition region, we first count the power of $\beta$ by substituting $k_p = \beta$ to (\ref{eq:ratiop}). Then,  from the Eqs.~(\ref{eq:Gf0}) and (\ref{eq:cE0i}), we  obtain 
\bea
\Omega_{\rm GW}|_{k_p = \beta} \sim \left( \frac{H_\star}{\beta} \right)^{- 2 \left( 1 + \frac{1}{p -1} + \frac{1}{1 - \tilde p} \right)} \ .
\eea
However, none of the known examples listed in Table~\ref{examplepost} has $\tilde p < 1/3$.  Therefore, it is an open question whether this can be a realistic case. 
\item[${\bullet }$]
In the case of quasi-de Sitter inflation plus kination domination after inflation, we have $i_{\rm OSC} = 0$. Therefore, the profile in the oscillatory region is nearly flat as shown in Fig.~\ref{fig:dS_tp_RD_with_p} and the lower panel of Fig.~\ref{fig:MDRD}. 

\end{enumerate}

From Figs.~\ref{fig:dS_tp_RD_with_p}, \ref{fig:MDRD}, \ref{fig:RDwRDplot1} and \ref{fig:RDtp1RDplot}, we can see that all the GW spectrum curves coincide in the deep IR region. The reason is that the modes in the deep IR region are out-of-horizon when they are produced. As a result, the amplitudes of these modes does not red-shift until re-entering the horizon. However, for the examples shown in these plots, the wavelengths of these modes are long enough such that they all re-enter the horizon in the RD stage after the complicated intermediate stages. Hence,  the strength of the GW signal, $\Omega_{\rm GW}$, is determined by the ratio of the GW energy density and the energy density of the radiation at the moment the mode re-enters the horizon. We have
\bea
\Omega_{\rm GW} \sim \frac{k_p^2 h^2_k}{H^2} \ .
\eea
At the moment the mode re-enters the horizon we have $k_p = H$. Therefore,  $\Omega_{\rm GW}$ is solely determined by the GW amplitude, which is related to the details of the GW source. This explains why the curves are all coincident in the deep IR region. 

From Figs.~\ref{fig:dS_tp_RD_with_p}, \ref{fig:MDRD}, \ref{fig:RDwRDplot1} and \ref{fig:RDtp1RDplot}, we can also see that in the case of a KD intermediate stage the GW signal is larger than that of the instantaneous reheating scenario. At the same time, it is still smaller than that of the scenario with an MD intermediate stage. This phenomenon can be easily understood since once the energy energy is stored in the form of the kinetic energy, it red-shifts as $a^{-6}$, which is much faster than radiation. As a result, the relative ratio of the GW energy density to the total energy density of the universe becomes larger, and therefore we get a larger $\Omega_{\rm GW}$. Whereas in the case with an MD intermediate stage, the total energy density red-shifts much slower ($a^{-3}$), resulting in a smaller $\Omega_{\rm GW}$.

\subsection{Observed GW Signal frequency}

The GW frequency will be red-shifted from the frequency $f_\star$ when phase transition takes place to the current value
\begin{align}\label{eq:ftoday}
    f_{\text {today }}={f_{\star}}\times \frac{a\left(\tau_{\star}\right)}{a_{r}}  \left(\frac{g_{* S}^{(0)}}{g_{* S}^{(R)}}\right)^{1 / 3} \frac{T_{\mathrm{CMB}}}{\left[\left(\frac{30}{g_{*}^{(R)} \pi^{2}}\right)\left(\frac{3 H_{r}^{2}}{8 \pi G_{N}}\right)\right]^{1 / 4}}~,
\end{align}
where $g_{* S}^{(0)}\simeq 3.91$ is the present value of the effective number of relativistic species. We also take $g_{*S}^{( {R})}\simeq g_{*}^{( {R})}\simeq 100$ to be the effective number of relativistic species at RD. The CMB temperature is taken as $T_{\rm CMB} \simeq 2.72$ K. 

Based on our discussion of the spectral shapes,  the GW signal peaks around $k_p = H_\star$. From Eq.~(\ref{eq:ftoday}),  the peak frequency today is red shifted to
\beq
\tilde{f}^{\rm peak}_{\text{today}} = 1.1 \times 10^{11} \text{Hz}  \times \left( \dfrac{H_{\rm end}}{m_\text{pl}} \right)^{1/2}  \left( \dfrac{a_r}{a_{\rm end}} \right)^{-\frac{1}{2 \tilde{\alpha} -1} - \frac{1}{2} }   \frac{a\left(\tau_\star\right)}{a_{\rm end}} ~.
\eeq

\begin{figure}[h!]
\centering
\includegraphics[width=15cm]{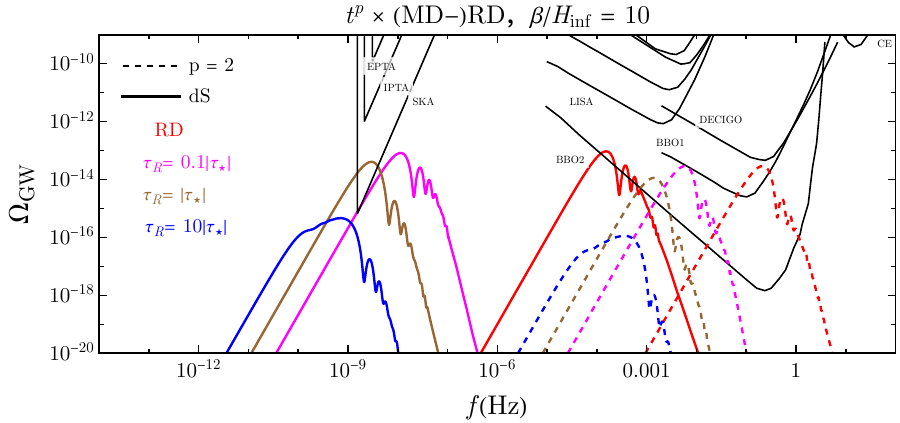}
\caption{The GW spectra from the phase transition for $t^p$ inflation and (MD-)RD post-inflationary evolution with $H_{\rm inf} \simeq H_\star = 10^8$GeV. The solid line is plotted using $p \rightarrow \infty$ (dS inflation) and $a_\star/a_{\rm end}= \exp(-19)$ while the dashed line is for $p = 2$ and $a_\star/a_{\rm end} = \exp(-23)$. Four different colors correspond to different $\tau_R$ settings. We set $\beta/H_\star$ to be 10. The overall strength of the GW signal scales linearly with $ {\Delta \rho_{\rm vac}}/{\rho_{\rm inf \star}} $.  For this plot, we have taken the conservative choice of $ {\Delta \rho_{\rm vac}}/{\rho_{\rm inf \star}} =0.1$. The curve for the sensitivity of BBO phase 2 (BBO2) is from \cite{Harry:BBO2}. Curves for sensitivities of other detectors are from \cite{Moore:2014lga}.  }
\label{fig:signal-tp-MDRD}
\end{figure}

\subsection{Observed GW signal in various cosmological scenarios}

Here we present numerical results of some examples of today's GW spectrum produced by first-order phase transition duting inflation. 

\subsubsection{$t^p \times$ MD-RD}

In Fig~\ref{fig:signal-tp-MDRD}, we show the explicit GW spectrum as a function of today's frequency generated by first-order phase transition during inflation, in which the solid curves are for quasi-de Sitter inflation and dashed curves for $t^p$ inflation with $p=2$. For the phase transition, we fix $H_\star = 10^8$ GeV during the phase transition. We also fix $\beta/H_\star = 20$. For the quasi-de Sitter inflation, we assume the phase transition happened at 19 e-folds before the end of inflation and for the $t^p$ scenario we assume the phase transition happened at $23$ e-folds before inflation. 
We assume a intermediate MD stage before reheating. To illustrate the effect of the intermediate MD stage, choose $\tau_R / |\tau_\star|$ = 0.1, 1 and 10. For comparison, we also show the instantaneous reheating scenario. According to the discussions in Sec.~\ref{sec:strength}, we know that if we plot the GW strength as a function of the physical wave number during the phase transition, the deep IR parts of the spectrum are coincident as shown in Fig.~\ref{fig:dS_tp_RD_with_p}, \ref{fig:MDRD}, \ref{fig:RDwRDplot1} and \ref{fig:RDtp1RDplot}. During MD the total energy density red-shifts slower than during RD, therefore more expansion of the universe is needed to cool it to today's temperature. As a result, with an extended MD intermediate stage after inflation, today's GW spectrum will be red-shifted to a lower frequency than in the instantaneous reheating scenario, as shown in Fig.~\ref{fig:signal-tp-MDRD}. For the same reason, the strength of the GW signal will get more suppressed with longer during of the MD stage.

\subsubsection{KD intermediate stage, dS $\times t^{\tilde{p} = 1/3}$-RD}

\begin{figure}[h!]
\centering
\includegraphics[width=14cm]{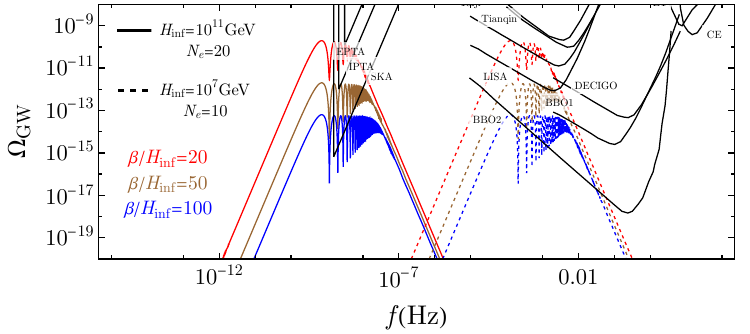}
\caption{The GW spectrum $\Omega_{\rm GW}$ in the dS $\times t^{\tilde{p} = 1/3}$-RD scenarios for different parameter sets. The solid line represents $\beta/H_{\rm inf} = 10$ and the dashed line is for $\beta/H_{\rm inf} = 50$. The red line has parameters $H_{\rm inf}=10^7 $ GeV, $N_e=8$ and $N_K=9$ and  the brown color has parameters  $H_{\rm inf}=10^{12} $ GeV, $N_e=15$ and $N_K=11$ respectively.} \label{fig:dS_tp-RD}
\end{figure}

In the case that the universe undergoes a KD stage after the end of inflation, the equation of state is $w =1$, which is the maximal value allowed by causality~\cite{Gouttenoire:2021jhk}. In this case, the dilution of the energy density goes like $\rho \sim a^{-6}$. Therefore, compared to the instantaneous reheating scenario, the relative GW signal strength is enhanced. In Fig.~\ref{fig:dS_tp-RD},  we show today's GW spectrum in the KD intermediate stage scenario. Here we assume a de Sitter inflation, with different choices of $\beta/H$. From the app.~\ref{app:tp-RD}, during the kination stage, we have for $\tau \gg |\tau_{\rm end}|$
\begin{equation}
    \dfrac{k}{a H} = 2 k \tau \ , \quad\quad \dfrac{a(\tau)}{a_{\rm end}} = \brr{\dfrac{2 \tau}{|\tau_{\rm end}|}}^{1/2} \ , \quad\quad \dfrac{H(\tau)}{H_{\rm end}} = \brr{\dfrac{|\tau_{\rm end}|}{2 \tau}}^{3/2}\ .
\end{equation}
The factor which controls the evolution of the GW signal is 
\begin{equation}
    \dfrac{1}{a(\tau)^4 H(\tau)^2} = \dfrac{2 |\tau_{\rm end}| \tau}{a_{\rm end}^2}  
\end{equation}

We denote $\tau_R$ as the conformal time when the universe enters RD stage or starts to reheat. Similarly we can define the e-fold $N_K$ characterizing the duration kination dominated stage. Its relation with the conformal time $\tau_R$ is given by
\begin{equation}
    N_K = \log \dfrac{a_R}{a_{\rm end}}=\dfrac{N_e + \log 2}{2}  + \dfrac{1}{2}\log\dfrac{\tau_R}{|\tau_\star|}
\end{equation}
The constraint on $N_K$ from BBN is studied in Ref.~\cite{Gouttenoire:2021jhk}.  
Fig.~\ref{fig:dS_tp-RD} presents the GW spectra with different parameters shown in the legend. In particular, we choose the $N_e = 8,N_K = 9$ and $N_e=15,N_K=11$ for the red and brown curves, respectively. One can see that with the extra enhancement due to the KD era, the GW signal can be observed by the next generation of the space GW detector (i.e. LISA~\cite{Audley:2017drz}, Tianqin~\cite{Luo:2015ght} and Taiji~\cite{Guo:2018npi}).


\begin{figure}[h!]
\centering
\includegraphics[width=14cm]{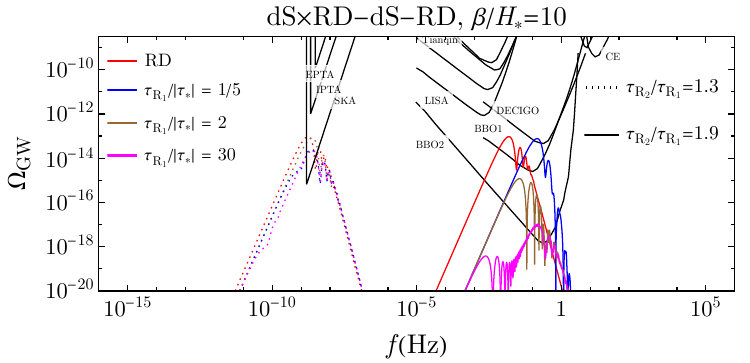}
\caption{The GW spectrum $\Omega_{\rm GW}$ in the dS $\times $RD-dS-RD scenarios for different parameter sets. We choose $\beta = 10 H_\tr{inf}, H_\tr{inf} = 10^{12}\tr{GeV}$. The solid line is plotted using $a_\star/a_{\rm end}=\exp(-22)$ and the dotted line is with $a_\star/a_{\rm end}=\exp(-38)$. $\tau_{R_1}/|\tau_\star| = (a_{R_1}/a_{\rm end})/(a_{\rm end}/a_\star)$ and $\tau_{R_2}/\tau_{R_1} = 2 - a_{R_1}/a_{R_2}$} \label{fig:Omega_dS-RD-dS-RD}
\end{figure}

\subsubsection{dS $\times$ RD-dS-RD}
As another interesting extension, which also shows very different signal pattern, we consider the scenario with a second period of dS-like inflation, occurring in the middle of two RD periods. The factor $\cE^i_0(k)$ has  been calculated in Sec.~\ref{subsec:osc}. For the scenario considered here, we have
\begin{eqnarray}
\left[ \cE^i_0(k) \cG^f_0(k)\right]^2 \frac{H_\star^2}{H_r^2} \left(  \frac{a_\star}{a_r} \right)^4 &=& \left( \dfrac{H_{\rm end}}{k}\right)^2  \cE^i_0(k)^2 \left( \dfrac{a_\star}{a_{\rm end}}\right)^4 \left( \dfrac{a_{R_1}}{a_r}\right)^4  \nn 
&=&  \left(  \dfrac{a_\star H_{\rm end}}{k} \right)^4  F\left(  \dfrac{k}{a_\star  H_{\rm end}}, \dfrac{\tau_{R_1}}{|\tau_\star|}, \dfrac{\tau_R}{\tau_{R_1}} \right)
\end{eqnarray}
in this case we have $\tau_R = \tau_{R_2}$. In the last equation we have inserted the explicit expressions and write the final result in terms of the dS-RD result and the ratio which is  only dependent on $k/(a_\star H_{\rm end})$ or $k |\tau_\star|$, the ratios of $\tau_{R_1}$ and $\tau_R$ to $|\tau_\star|$ but not $|\tau_{\rm end}|$. Of course while $\tau_{R_2} \rightarrow \tau_{R_1}$, the ratio $F \rightarrow 1$ and we recover the simple dS-RD result. In Fig.~\ref{fig:Omega_dS-RD-dS-RD} we show the observed GW spectrum under two different $\tau_{R_2}/\tau_{R_1}$ settings and the corresponding four $\tau_{R_1}$ values. One can see that if $\tau_{R_1} \gg |\tau_{\rm end}|$ the frequency redshift is independent of $\tau_{R_1}/|\tau_\star|$ and only depends on the ratio $\tau_{R_1}/\tau_{R_2}$. The intermediate dS stage does not alter the total energy density or Hubble parameter but only makes the scale factor increase. 
Compared to scenario with a single RD period following the reheating (solid red curve in the picture), the large $\tau_{R_1}/|\tau_\star|$ leads to a suppression of the signal strength except the deep IR modes. The deep IR modes only re-enter the horizon after $\tau_{R_2}$. The deep UV modes re-enter the horizon at the first RD and never exit the horizon during the intermediate dS stage, receiving additional dilution from the extra dS expansion. Some of the intermediate modes exit the horizon again at the inserted dS stage. This can produce additional, although suppressed, oscillatory pattern. The details can be found in Sec.~\ref{app:RD_tp_RD}. 

The modes re-entering the horizon in the RD stage between the two dS stages may exit the horizon again in the second dS stage and produce addition oscillatory patterns in the spectrum. This is shown in Fig.~\ref{app:RD_tp_RD} where the additional wiggles in the oscillatory region of the spectrum are visible. Clearly, the additional oscillation in the spectrum is due to the second dS stage since it makes the Hubble radius smaller and some modes which have entered the horizon earlier become super-horizon again. We can also find extra oscillatory behavior with $k$ analogue in the first dS stage. Some modes in the IR region with relatively large momentum, which was already out of the horizon when produced, are inside the horizon of the second dS stage. These modes can now exit the horizon in the second dS stage. This is clearly shown in the IR part of the dashed curves in Fig.~\ref{app:RD_tp_RD}, where one can see small periodic wiggles on top of the $k^3$ slopes.

\section{Summary and future directions}
\label{sec:summary}

The large field distance the inflaton may traverse during inflation can trigger interesting dynamics in a spectator sector. This can be the case even if the couplings between the inflaton and the spectator sectors are suppressed by some high energy scales.  As an example, we consider a first-order phase transition in the spectator sector during the inflation and its associated GW signal. 
Such an event can be treated approximately as an instantaneous source of GW. In Ref.~\cite{An:2020fff}, we pointed out that there is an oscillatory feature in the spectrum of the GW signal. In this paper, we offer a more detailed discussion of this feature and the spectrum. In addition to a first-order phase transition, this feature can also arise for any other instantaneous source. Its discovery can be an unmistakable signal of such a dramatic event during inflation.

The cosmological background the GWs propagated through would leave imprints on the final GW spectrum we observe today. As such, the shape of the GW signal will also offer a new window on the cosmological evolution in the early universe.  
 In particular, we have demonstrated that if there is no second inflationary stage during the RD, the IR part of the GW spectrum is only sensitive to post-inflationary history. The UV and intermediate part of the GW spectrum are, however, sensitive to both the post-inflationary history and the inflationary scenario. Thus the stochastic GW background offers a promising  and complementary new probe in probing the early universe histories. This is particularly interesting for the epochs close to the end of the inflation and before the BBN, which can not be directly probed by CMB, large scale structure, and other cosmological observables. In this paper, we consider several examples, including different expansion stages either during the inflation or after the reheating. Using the spectral information in the GW signal, we can clearly distinguish these scenarios. 

There are a lot of new directions to pursue further. In addition to a strong first-order phase transition, there could also be additional mechanisms to generate an approximate instantaneous source of the GW. In these cases, the GW signal would have the same feature as discussed in this paper. It would be interesting to consider these mechanisms in detail and study the feasibility of observing such signals. 
Instantaneous sources during inflation will not only induce an oscillatory feature on the GW spectrum but also on the scalar perturbations. If such oscillatory features are seen on both tensor and scalar perturbations, that will be a strong hint that it is induced by first-order phase transition during inflation.
The large inflaton field excursion can also trigger many different dynamics in the spectator sector, such as the production of non-perturbative objects, confinement, etc. 
We leave detailed studies of these directions to future works.

\section*{Acknowledgments}
We would like to thank Jeff Dror, Junwu Huang, Wayne Hu, Austin Joyce, Keisuke Inomata, Soubhik Kumar, Hayden Lee, Subodh P. Patil, Tomislav Prokopec, Xi Tong, Andrea Tesi, Dong-Gang Wang and Chen Yang for discussions. LTW would like to thank the hospitality of Galileo Galilei Institute. HA is supported in part by the National Key R\&D Program of China under Grant No. 2021YFC2203100 and 2017YFA0402204, the NSFC under Grant No. 11975134, and the Tsinghua University Initiative Scientific Research Program. KFL is partially supported by the DOE grant  DE-SC0022345. LTW is supported by the DOE grant DE-SC0013642. The work of SZ is supported by in part by JSPS KAKENHI Grant Number 21F21026. 

\appendix

\section{Details of the first order phase transition models during inflation and the estimation of $\beta/H$}\label{model}
\begin{figure}[t] 
	\centering 
	\includegraphics[width=7cm]{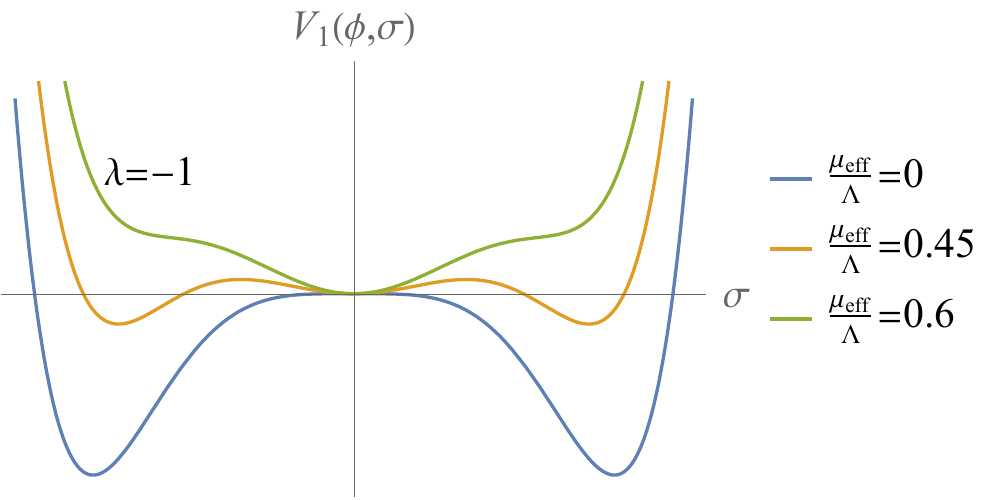} 
	\includegraphics[width=7cm]{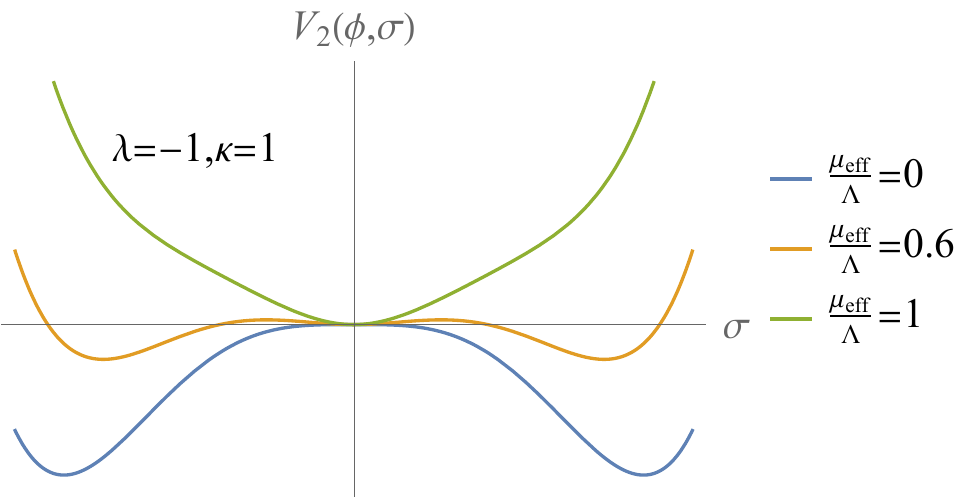} 
	\includegraphics[width=7cm]{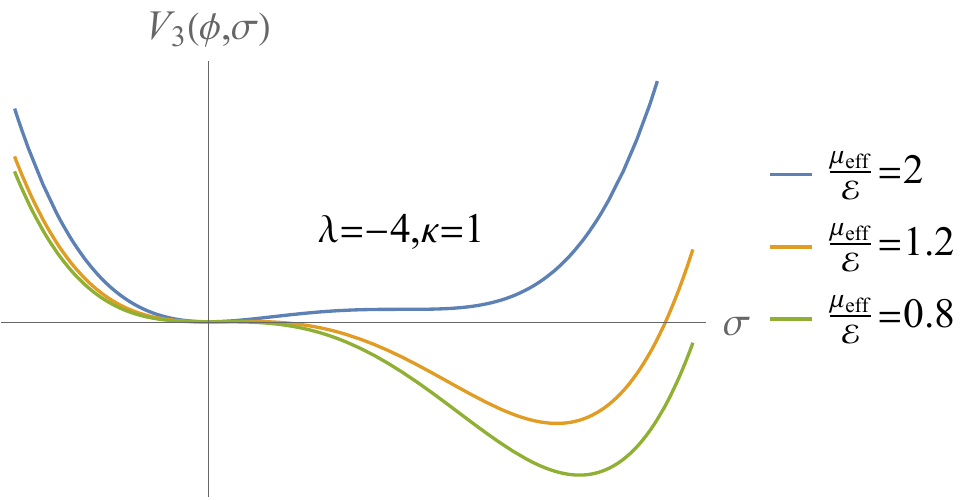}  
	\caption{This figure depicts the three typical potentials of the $\phi$ and $\sigma$ field $V_1(\phi,\sigma)$, $V_2(\phi,\sigma)$ and $V_3(\phi,\sigma)$.
	} \label{potential}
\end{figure}

In this section, we give a brief summary of the details of the model of first order phase transition. We consider three concrete models for illustration purposes. 
\begin{align}
	&V_{1}(\phi, \sigma)=-\frac{1}{2}\left(\mu^{2}-c^{2} \phi^{2}\right) \sigma^{2}+\frac{\lambda}{4} \sigma^{4}+\frac{1}{8 \Lambda^{2}} \sigma^{6}~, \\
	&V_{2}(\phi, \sigma)=-\frac{1}{2}\left(\mu^{2}-c^{2} \phi^{2}\right) \sigma^{2}+\frac{\lambda}{4} \sigma^{4}+\frac{\kappa}{4} \sigma^{4} \log \frac{\sigma^{2}}{\Lambda^{2}}~, \\
	&V_{3}(\phi, \sigma)=-\frac{1}{2}\left(\mu^{2}-c^{2} \phi^{2}\right) \sigma^{2}+\frac{\lambda}{3} \mathcal{E} \sigma^{3}+\frac{\kappa}{4} \sigma^{4}~.
\end{align}
During inflation, the inflaton field value decreases. As a result, the effective mass of the $\sigma$ field $\mu_{\rm eff}^2 \equiv -(\mu^2-c^2\phi^2)$ evolves from a positive value to a negative value. Thus the $\sigma$ potential goes from a symmetric phase to a symmetry broken phase. For certain parameters, the phase transitions in these models can be first-order, as shown in Fig.~\ref{potential}. 

Next, we consider typical value of $\beta$ in the models. 
\begin{align}\nonumber
	\beta & = \bigg| \frac{dS_4}{dt} \bigg| = \bigg| \frac{dS_4}{d\mu_{\rm eff}} \bigg|\bigg| \frac{d\mu_{\rm eff}}{dt} \bigg| \\
	& = \bigg| \frac{dS_4}{d\log\mu_{\rm eff}^2} \bigg| \bigg| \frac{d\mu_{\rm eff}^2}{\mu_{\rm eff}^2 dt} \bigg|=I_1S_4\bigg|\frac{2\dot \phi}{\phi-\frac{\mu^2}{c^2\phi}} \bigg|~.
\end{align}
where 
\begin{align}
	I_1 = \frac{1}{S_4}\bigg|\frac{dS_4}{d\log\mu_{\rm eff}}\bigg|~.
\end{align}
The typical values of $S_4$ and $dS_4/d\log {\mu_{\rm eff}}$ can be computed numerically using CosmoTransitions~\cite{Wainwright:2011kj}. Our results show that the value of $I_1$ is typically $0.2\sim5$~\cite{An:2020fff}. 

The value of $S_4$ is determined by requiring the phase transition to complete during inflation, and in this model, it is determined to be 
\bea
S_4 \approx \log\left( \frac{m_\sigma^4}{\beta^4} \right) \ .
\eea
The details of the derivation is can be find in the appendix of Ref.~\cite{An:2020fff}. One can see that the typical value of $S_4$ is about ${\cal O}(100)$. 

In slow-roll inflation, we have the following relations
\begin{align}
	\int_{\phi_{\rm end}}^{\phi_{\rm PT}} \frac{d\phi}{\sqrt{2\epsilon}M_{\rm pl}} = N_{\rm e}~,
\end{align}
where $N_{\rm e}$ is the e-folding number before the end of inflation. Assuming the phase transition happened during Since we can always shift the value of the inflation potential such that $\phi_{\rm end} = 0$, we can use $\phi_{\rm PT}$ to estimate the value of $\phi$ at the phase transition time. 

Putting the above estimations together, we have an estimation for $\beta/H$ that
\begin{align}
	\frac{\beta}{H} \simeq I_1 S_4 \times \frac{1}{N_{\rm e} \big| 1- \frac{\mu^2}{c^2\phi_{\rm PT}^2} \big|}~.
\end{align}
During phase transition, the value of $c^2 \phi_{\rm PT}^2 - \mu^2$ changes from positive to negative. The value of $|1 - \frac{\mu^2}{c^2 \phi_{\rm PT}^2}|^{-1}$ can vary from ${\cal O}(1)$ to ${\cal O}(10^{-2})$, depending on the details of the parameters. Therefore, one can see that in this model, it is highly probable that the value of $\beta/H$ is around ${\cal O}(10)$ to ${\cal O}(100)$.

\section{Transition} \label{transition}
In this section, we provide the detailed calculation of $\tilde{\mathcal{G}}$ and $\tilde {\mathcal{E}}$ for various inflation and post-inflation scenarios.
 
\subsection{dS-$t^p$-dS inflation scenario}\label{inflationaryscenarios}

In this subsection, we study the consequences of the inflation scenario, in which the inflation era is composed by a quasi-de Sitter stage followed by a $t^p$ stage with $p>1$, and then followed by another quasi-de Sitter stage before reheating. There are many studies in the literature considering such a scenario of two-stage inflation with a break~ \cite{Pi:2019ihn}. 

The Green's function $\cG$ can be written as
\begin{equation}
\tilde{\mathcal{G}} (\eta, \eta') = \left\{
             \begin{array}{ll}
            & \left(\frac{1}{\eta+\frac{\eta_{\rm dS_2}-\eta_{\rm dS_1}}{p}}-\frac{1}{\eta^{\prime}}\right) \cos \left(\eta+\frac{\eta_{\rm dS_2}-\eta_{\rm dS_1}}{p}-\eta^{\prime}\right)\vspace{0.3cm} \\
            & +\left(1+\frac{1}{(\eta+\frac{\eta_{\rm dS_2}-\eta_{\rm dS_1}}{p}) \eta^{\prime}}\right) \sin \left(\eta+\frac{\eta_{\rm dS_2}-\eta_{\rm dS_1}}{p}-\eta^{\prime}\right)
              ,\quad   \eta < \eta_{\rm dS_1} \vspace{0.3cm}  \\
           &\sqrt{\eta + \dfrac{\eta_{\rm dS_1}}{ p - 1}} \bigg[   C_1  J_{3/2 + 1/(p - 1)} \brr{\eta + \dfrac{\eta_{\rm dS_1}}{p-1}}   \\ 
            &+  C_2 Y_{3/2 + 1/(p - 1)} \brr{\eta + \dfrac{\eta_{\rm dS_1}}{p-1}}   \bigg]
           ,\quad  \eta_{\rm dS_1} < \eta < \eta_{\rm dS_2} \vspace{0.3cm} \\
           &C_3  \srr{ \cos\brr{ \eta - \eta'}   - \dfrac{\sin\brr{ \eta  - \eta'}}{ \eta  } }+  
            C_4   \srr{ \sin\brr{ \eta  - \eta'} +  \dfrac{\cos\brr{ \eta  - \eta'}}{ \eta }} ,\quad   \eta > \eta_{\rm dS_2} 
             \end{array}
\right.
\end{equation}
where the coefficients $C_{3,4}$ can be determined straightforwardly using the evolution of the scale factor, given in Eq.~(\ref{eq:ataudStpdS}). However, the expressions are lengthy and are not presented here. 
Then from the above formula we can derive the function $\cG_0^f$ in this scenario,
\begin{equation}
(\tilde{\mathcal{G}}_0^f)^2 = \dfrac{H_{\rm dS_2}^2}{k^2} \brr{ C_3^2  +  C_4^2 }~.
\end{equation}
Inserting the $\cE^i_0$ for RD we can get the combination
\begin{equation}
\begin{split}
&\left[ \cE^i_0(k) \cG^f_0(k)\right]^2 \dfrac{H_\star^2}{H_r^2} \left(  \dfrac{a_\star}{a_r} \right)^4 =  \dfrac{H_{\rm dS_2}^2}{k^2} \brr{ C_3^2  +  C_4^2 } \brr{\dfrac{a_r^2 H_{\rm dS_2}}{k}}^2 \dfrac{H_{\rm dS_1}^2}{H_{\rm dS_2}^2} \brr{ \dfrac{a_\star}{a_r}  }^4 \\
=& \dfrac{H_{\rm dS_1}^2 H_{\rm dS_2}^2}{k^4} a_\star^4  \brr{ C_3^2  +  C_4^2 } = \brr{ \dfrac{H_{\rm dS_1} a_\star}{k} }^4  \brr{1 + \dfrac{p-1}{p} \dfrac{\tau_{\rm dS_2} - \tau_{\rm dS_1}}{\tau_{\rm dS_1}}}^{1/(p-1)}~ \brr{ C_3^2  +  C_4^2 }.
\end{split}
\end{equation} 

\subsection{Post-inflationary scenarios}\label{postinflationaryscenarios}
In this subsection, we discuss the detailed calculations in various post-inflationary scenarios. 

\subsubsection{$t^\tp$-RD}\label{app:tp-RD}
First, we consider the general power law expansion $t^\tp$-RD, before reheating with $\tp <1$. The scale factor can be parameterized as  
\begin{equation}
a = \left\{
             \begin{array}{lr}
             a_{\rm end} \brr{\tau - C_1 \tau_{\rm end}}^{\tilde{\omega}} \brr{\tau_{\rm end} - C_1 \tau_{\rm end}}^{-\tilde{\omega}}~, & \tau_{\rm end} < \tau < \tau_{R} \vspace{0.3cm}  \\
             a_{R} \srr{ 1 + a_{R} H_{R} \brr{\tau-\tau_{R}}}~,  &  \tau > \tau_{R}
             \end{array}
\right.
\end{equation}
where
\begin{equation}
    \tilde{\omega} = \tilde{p}/\brr{1-\tilde{p}} \quad 
    C_1 = 1- \dfrac{\tilde{\omega}}{a_{\rm end} H_{\rm end} \tau_{\rm end}} \quad a_R = a(\tau_{R}) \quad H_R = H(\tau_{R})
\end{equation}
During the $t^\tp$ expansion stage, we have
\begin{equation}
    \dfrac{a''}{a} = \dfrac{\tilde{\omega} \brr{\tilde{\omega} - 1}}{\brr{\tau - \tau_{\rm end} + \dfrac{\tomega}{a_{\rm end}H_{\rm end}}}^2} \ .
\end{equation}
Hence the solution to the Green's function is
\begin{align}
    a \mathcal E 
    = \left\{
           \begin{array}{lr}
            A_5  \sqrt{\eta + \eta_{R} (\tomega -1)} J_{\tomega-1/2} \srr{ \eta +  \eta_{R} (\tomega -1) } + & {} \\
             B_5 \sqrt{\eta + \eta_{R} (\tomega -1)} Y_{\tomega-1/2} \srr{ \eta + \eta_{R} (\tomega -1) } ~, &  \tau_{\rm end} < \tau < \tau_{R} \vspace{0.3cm}  \\
             A_6 \cos(\eta) + B_6 \sin(\eta) ~, &  \tau > \tau_{R}
             \end{array}
\right.
\end{align}
Imposing the initial condition $\mathcal{E}(\tau_{\rm end}) = 1$ and $\mathcal{E}'(\tau_{\rm end}) = 1$, we can get the value of $A_5$ and $B_5$. Then the continuity condition at $\tau_{R}$ can lead to explicit expressions of $A_6$ and $B_6$. For $\tau_R \gg |\tau_\star|$, we can simplify the result
\begin{equation}\label{eq:app_tp_RD_aH}
    \dfrac{a(\tau_R)}{a_{\rm end}}
    = \brr{a_{\rm end} H_{\rm end} \,\tau_{R}\, \tilde{\omega}^{-1} }^{\tilde{\omega}}
    \quad \dfrac{H(\tau_{R})}{H_{\rm end}} =   \brr{\dfrac{\tilde{\omega}}{a_{\rm end} H_{\rm end} \tau_{R}}  }^{\tilde{\omega} + 1} 
\end{equation}
Hence 
\begin{equation}
    \dfrac{k}{a_{R} H_{R}} = \dfrac{k \tau_{R}}{\tilde{\omega} } \quad\quad\quad \dfrac{1}{a_R^4 H_R^2} = \dfrac{1}{\tilde{\omega}^2 a_{\rm end}^2} \brr{\dfrac{\tilde{\omega}}{a_{\rm end} H_{\rm end}}}^{2\tilde{\omega}} \tau_{R}^{2-2\tilde{\omega}}
\end{equation}
We can see for $\tilde{\omega} < 1$, the latter factor is a growing function with increasing $\tau_R$. The critical value is $\tilde{\omega}=1$ or $\tilde{p} = 1/2$ which is just RD. For example the kination dominated stage gives $\tilde{p} = 1/3$ can give more enhanced GW signal. The expressions of $A_6$ and $B_6$ are easy to derive but complicated, and therefore are not presented here.

\subsubsection{MD-RD}\label{app:MDRD}
In this subsection, 
We discuss a specific case that the intermediate stage before reheating is MD. 
In this case, the scale factor can be written as
\bea
a = \left\{
             \begin{array}{lr}
             a_{\rm end} \srr{1+\dfrac{1}{2} a_{\rm end} H_{\rm end} (\tau - \tau_{\rm end})}^2~, & \tau_{\rm end} < \tau < \tau_{\rm MR} \vspace{0.3cm}  \\
             a_{\rm MR} + \dfrac{8 a_{\rm MR}^2}{a_{\rm end}^3 H_{\rm end}^2 \tau_{\rm MR}^3} (\tau - \tau_{\rm MR})~,  &  \tau > \tau_{\rm MR}
             \end{array}
\right.
\eea
where $a_{\rm end}$ and $a_{\rm MR}$ is the abbreviation of $a(\tau_{\rm end})$ and $a(\tau_{\rm MR})$. $H_{\rm end}$ is the Hubble value during inflation. Here we have used the condition that $a$ is continuous at $\tau_{\rm MR}$, so $a_{\rm MR}=a_{\rm end}^3 H_{\rm end}^2 \tau_{\rm end}^2/4$ and $H_{\rm MR} = 8/a_{\rm end}^3 H_{\rm end}^2 \tau_{\rm MR}^3$. The approximation $\tau_{\rm MR}\gg |\tau_{\rm end}|$ is exploited. The corresponding $a''/a$ is
\begin{equation}
\dfrac{a''}{a} = \left\{
             \begin{array}{lr}
             \dfrac{a_{\rm end}^2 H_{\rm end}^2}{2\brr{1 + \frac{1}{2} a_{\rm end} H_{\rm end} ( \tau - \tau_{\rm end})}^2} = \dfrac{2}{(\tau-\tau_{\rm end} + \frac{2}{a_{\rm end} H_{\rm end}})^2}~ , & \tau_{\rm end} < \tau < \tau_{\rm MR} \vspace{0.3cm}  \\
           0~, &  \tau > \tau_{\rm MR}
             \end{array}
\right.
\end{equation}
The $\mathcal{E}(\eta)$ in MD stage must be proportional to the product of  the Green's function and the inverse scale factor, namely
\beq
\mathcal{E}(\eta) \sim \dfrac{k \tilde{G} }{a(\tau)} \sim \dfrac{k \tilde{G}}{\srr{1+\dfrac{a_{\rm end} H}{2k} (\eta - \eta_{\rm end})}^2}~.
\eeq
For $k \tilde{G}$, we can just do a shift of the conformal time. The general solution is
\beq\label{eq:MD_GF}
k \tilde{G} = A_2 \brr{\cos\eta - \dfrac{\sin\eta}{\eta -  \eta_{\rm end} + \dfrac{2 k}{a_{\rm end} H_{\rm end}} }} + B_2 \brr{\sin\eta + \dfrac{\cos\eta}{\eta-\eta_{\rm end} + \dfrac{2 k}{a_{\rm end} H_{\rm end}}}}~.
\eeq
Imposing the initial condition at $\tau_{\rm end}$ we get
\beq
\begin{split}
A_2 &=\brr{1 - \dfrac{3 a_{\rm end}^2 H_{\rm end}^2}{4 k^2}  } \cos\eta_{\rm end}  -   \dfrac{3 a_{\rm end} H_{\rm end}}{2 k}\sin \eta_{\rm end}~, \\
B_2 &= \dfrac{3 a_{\rm end} H_{\rm end}}{2 k} \cos\eta_{\rm end}  +  \brr{1 -  \dfrac{3 a_{\rm end}^2 H_{\rm end}^2}{4 k^2}  } \sin\eta_{\rm end}~.
\end{split}
\eeq
At $\tau_{\rm end}$, we have $2 k/(a_{\rm end} H_{\rm end})  \ll 1$. Hence we can get $\mathcal{E}(\eta)$ in MD stage
\begin{equation}
\mathcal{E}(\eta) = \dfrac{3 a_{\rm end}^2 H_{\rm end}^2}{4 k^2} \brr{ \dfrac{\sin\eta}{\eta}  - \cos\eta   } \dfrac{a_{\rm end}}{a(\tau)}  \xrightarrow{ \tau \gg |\tau_{\rm end}|} \dfrac{3}{\eta^2} \brr{ \dfrac{\sin\eta}{\eta}  - \cos\eta   }~. \\
\end{equation}
If in the deep MD we have
\beq
\tilde{\mathcal{E}}^i_0 = \dfrac{3 a_{\rm end}^3 H_{\rm end}^2}{4 k^2}~.
\eeq
For the next RD stage, $\mathcal{E}$ can  be written as
\beq
\mathcal{E}(\eta) = \dfrac{A_3 \cos\eta + B_3 \sin\eta}{a(\tau)}~.
\eeq
Matching at $\tau_{\rm MR}$, we can solve the coefficients
\beq
A_3 = -\dfrac{3 a_{\rm end}^3 H_{\rm end}^2 \brr{\eta_{\rm MR}^2 -\sin^2 \eta_{\rm MR}} }{4 k^2 \eta_{\rm MR}^2}~, \quad\quad 
B_3 = \dfrac{3 a_{\rm end}^3 H_{\rm end}^2 \brr{\eta_{\rm MR} -\sin \eta_{\rm MR} \cos \eta_{\rm MR}} }{4 k^2 \eta_{\rm MR}^2}~.
\eeq
Hence we can get at subsequent RD
\beq
\tilde{\mathcal{E}}^i_0 =\sqrt{A_3^2 + B_3^2} = \dfrac{3 a_{\rm end}^3 H^2}{8 k^2 \eta_{\rm MR}^2} \sqrt{2\srr{ 1 + 2 \eta_{\rm MR}^4 + (2 \eta_{\rm MR}^2 -1)\cos(2\eta_{\rm MR}) - 2 \eta_{\rm MR} \sin(2 \eta_{\rm MR})  }}~.
\eeq

\subsubsection{RD-$t^\tp$-RD}\label{RDWRD}
\label{app:RD_tp_RD}

Here we present the useful formulae in the scenario of RD-$t^\tp$-RD. 
As we already see from~\ref{sec:IRGW}, the low frequency spectrum of the GW is independent of the details governing the generation of the GW. Instead, they are fixed by causality. They are only dependent on the evolution of the universe when the modes reenter the horizon~\cite{Hook:2020phx}.
In this case, the scale factor can be written as
\bea
a = \left\{
             \begin{array}{lr}
             a_{\rm end}  \srr{1 + C_1(\tau - \tau_{\rm end} )} ~ , & \tau_{\rm end} < \tau < \tau_{R_1} \vspace{0.3cm}  \\
             \dfrac{a_{R_1}}{\tau_{R_1}^\omega(1-C_2)^{\tomega}} \brr{\tau - C_2 \tau_{R_1}}^{\tomega}~, &  \tau_{R_1} < \tau < \tau_{R_2}  \vspace{0.3cm}\\  
             a_{R_2} \srr{ 1 + C_3 \brr{\tau - \tau_{R_2}}}~. &  \tau > \tau_{R_2}
             \end{array}
\right.
\eea
We choose the conformal time $\tau_{\rm end},\tau_{R_1},\tau_{R_2}$ and $a_{\rm end}, H_{\rm end}$ as free parameters and denote $\tomega \equiv \tp/(1-\tp)$. The scale factor $a$ and its first-order derivative are continuous at $\tau_{R_1}$ and $\tau_{R_2}$. In the calculation, we assume $\tau_{R_1}\gg |\tau_{\rm end}|$ to simplify the calculation. Then, under this limit, we have
\begin{equation}\nonumber
\begin{split}
C_1 &= H_{\rm end} a_{\rm end} \quad  a_{R_1} = a_{\rm end}^2 H_{\rm end} \tau_{R_1} \quad C_2 = 1-\tomega \\
a_{R_2} &= a_{R_1} \tau_{R_1}^{-\tomega} \srr{ \tau_{R_2} - \tau_{R_1} (1-\tomega) }^{\tomega}  \tomega^{-\tomega} \quad C_3 = \frac{\tomega}{-\tau_{R_1} + \tau_{R_2} + \tau_{R_1} \tomega }~.
\end{split}
\end{equation}
The corresponding $a''/a$ is
\begin{equation}
\dfrac{a''}{a} = \left\{
             \begin{array}{lr}
             0~, & \tau_{\rm end} < \tau < \tau_{R_1} \vspace{0.3cm}  \\
             \dfrac{(-1 + \tomega)\tomega}{ (\tau - C_2 \tau_{R_1})^2}~, &  \tau_{R_1} < \tau < \tau_{R_2} \vspace{0.3cm}  \\
             0~, &  \tau > \tau_{R_2}
             \end{array}
\right.
\end{equation}
The solution to the Green's function after $\tau_{\rm end}$ is then 
\begin{align}
    a \mathcal E 
    = \left\{
             \begin{array}{lr} 
             A_4 \cos(\eta) + B_4 \sin(\eta) ~, & \tau_{\rm end} < \tau < \tau_{R_1} \vspace{0.3cm}  \\ 
            A_5  \sqrt{\eta + \eta_{R_1} (\tomega -1)} J_{\tomega-1/2} \srr{ \eta +  \eta_{R_1} (\tomega -1) } + & {} \\
             B_5 \sqrt{\eta + \eta_{R_1} (\tomega -1)} Y_{\tomega-1/2} \srr{ \eta + \eta_{R_1} (\tomega -1) } ~, &  \tau_{R_1} < \tau < \tau_{R_2} \vspace{0.3cm}  \\
             A_6 \cos(\eta) + B_6 \sin(\eta) ~, &  \tau > \tau_{R_2}
             \end{array}
\right.
\end{align}
Using the fact that $\mathcal{E} = \sin\eta/\eta$ at $\tau_{R_1}$, we can see that $A_4=0$ and $B_4=C_1 a_{\rm end}/k$. Subsequently, we can match the first radiation dominated era with the era with general equation of state and get the corresponding coefficients $A_5$ and $B_5$
\beq
\begin{split}
A_5 &= -\dfrac{a_{\rm end}^2 H_{\rm end} \pi \sqrt{\tomega \eta_{R_1}} }{2 k } \srr{ Y_{\tomega-1/2} \brr{ \eta_{R_1} \tomega } \brr{\cos\eta_{R_1} - \dfrac{\sin\eta_{R_1}}{\eta_{R_1}}}  + Y_{\tomega+1/2} \brr{ \eta_{R_1} \omega } \sin\eta_{R_1} }~,\\
B_5 &= -\dfrac{a_{\rm end}^2 H_{\rm end} \pi \sqrt{\tomega \eta_{R_1}} }{2 k } \srr{ J_{\tomega-1/2} \brr{ \eta_{R_1} \tomega } \brr{\cos\eta_{R_1} - \dfrac{\sin\eta_{R_1}}{\eta_{R_1}}}  + J_{\tomega+1/2} \brr{ \eta_{R_1} \tomega } \sin\eta_{R_1} }~.
\end{split}
\eeq
The complete expression for $A_6$ and $B_6$ are lengthy, but the calculation is straightforward, therefore we do not list them here. We present the numerical result in the main text.

\subsubsection{RD-$t^{\tilde{p}=1}$-RD}\label{sec:RDtp1RD}
The typical RD-$t^\tp$-RD formulas do not apply for the case $\tp = 1$ cause $\tilde{\omega} = \infty$. For $\tp=1$, the Hubble sphere keeps invariant and the modes would stay within or out of horizon until this stage terminates. During $\tau_{R_1}$ and $\tau_{R_2}$ the scale factor is given by
\begin{equation}
    a = a_{R_1} e^{a_{R_1} H_{R_1} \brr{\tau - \tau_{R_2}}} \quad\quad \rightarrow \dfrac{a''}{a} = \dfrac{1}{\tau_{R_1}^2}
\end{equation}
The Hubble parameter is
\begin{equation}
    H(\tau) = \dfrac{H_{R_1}}{a/a_{R_1}}
\end{equation}
Hence depending the value of $k\tau_{R_1}$ during $t^{\tilde{p}=1}$ stage the solution of $a \mathcal{E}$ is a piecewise function
\begin{align}
    a \mathcal E 
    = \left\{
             \begin{array}{lr} 
             A_{41} \exp\brr{ \dfrac{k \tau \sqrt{1-\eta_2^2}}{\eta_2}  } + B_{41} \exp\brr{ -\dfrac{k \tau \sqrt{1-\eta_2^2}}{\eta_2}  } ~, &  \eta_2 < 1 \vspace{0.3cm}  \\ 
            A_{42}\, \tau + A_{52} & \eta_2 = 1\vspace{0.3cm} \\ 
            A_{43} \cos\brr{\dfrac{k \tau \sqrt{\eta_2^2-1}}{\eta_2}} + B_{53} \sin\brr{\dfrac{k \tau \sqrt{\eta_2^2-1}}{\eta_2}}  , &  \eta_2 > 1 
             \end{array}
\right.
\end{align}
in which $\eta_2 = k \tau_{R_1} = k/(a_{R_1} H_{R_1})$. Imposing the continuity condition on $\tau_{R_2}$ we can get the $a \mathcal{E}$ at $\tau > \tau_{R_2}$.

\subsection{RD-dS-RD}\label{app:RD-dS-RD}

Here we present the detailed discussions of the RD-dS-RD scenario. 
This is the special case with $\tomega=-1$ in Section~\ref{app:MDRD}. The evolution of the GW after inflation can be parameterized as   
\begin{equation}\label{eq:A27}
    a \mathcal E = \left\{
             \begin{array}{ll} 
             A_4 \cos(\eta) + B_4 \sin(\eta) ~, & \tau_{\rm end} < \tau < \tau_{R_1} \vspace{0.3cm}  \\
             A_5  \srr{ \cos(\eta - 2 \eta_{R_1}) - \dfrac{\sin(\eta - 2 \eta_{R_1})}{  \eta  -  2 \eta_{R_1}  }  } + \vspace{0.2cm}\\   B_5  \srr{\sin(\eta - 2 \eta_{R_1} ) + \dfrac{\cos(\eta - 2 \eta_{R_1}) }{ \eta - 2 \eta_{R_1} }   } ~, &  \tau_{R_1} < \tau < \tau_{R_2} \vspace{0.3cm}  \\
             A_6 \cos(\eta) + B_6 \sin(\eta) ~, &  \tau > \tau_{R_2}
             \end{array}
\right.
\end{equation}


Using the fact that $\mathcal{E} = \sin\eta/\eta$ at  $\tau_{R_1}$, we can match the subsequent stages and get the coefficients. Now $\tilde{\mathcal{E}}_i^0$ is equal to $(A_6^2 + B_6^2)^{1/2}$. Obviously $\tau_{R_1}<\tau_{R_2} < 2 \tau_{R_1}$. The Hubble parameter at $\tau_{R_1} < \tau < \tau_{R_2}$ is $H_{R_1} = (a_{R_1} \tau_{R_1})^{-1}$. Hence during this dS stage, we have
\begin{equation}
\dfrac{k}{a H} = k (2 \tau_{R_1} -\tau)~.
\end{equation}
To illustrate the distinction between this case with no intermediate dS space. The two scenarios have the same initial condition, namely the phase transition take place at same conformal time and scale factor $a(\tau_\star)$ and the later evolution during the first inflation is identical. What is more, we require the same function $H(a)$ once entering the last RD stage so that the observatory condition today is same for the two scenarios. To discuss the effect of the intermediate dS stage, We define the quantity  
\begin{equation}\label{eq:def_R_RD-dS-RD}
R = (\tilde{\mathcal{E}}_i^0)_{\tr {int}} / (\tilde{\mathcal{E}}_i^0)_{\tr {RD}} ~,  
\end{equation}
where $(\cE^0_i)_{\rm RD}$ is the function $\cE^0_i$ in the instantaneous reheating scenario given in Eq.~(\ref{eq:3.21}) and $(\cE^0_i)_{\rm int}$ is $\cE^0_i$ calculated in RD-dS-RD scenario and can be derived from Eq.~(\ref{eq:A27}).

\begin{figure}[t]
\centering
\includegraphics[height=5.1cm]{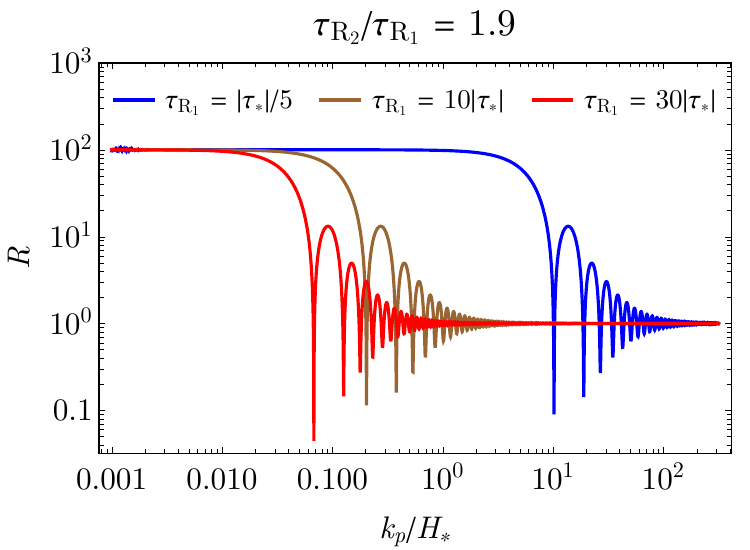}
\includegraphics[height=5.1cm]{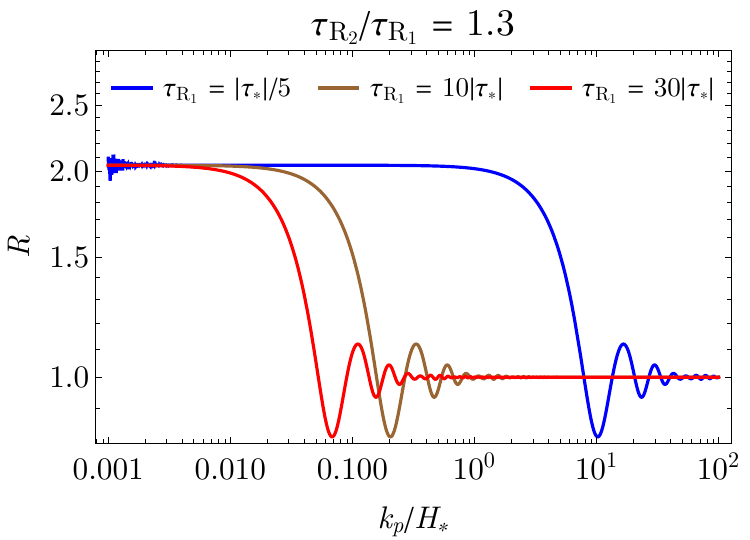}
\caption{The ratio R of $\tilde{\mathcal{E}}_i^0$ in dS-RD and single RD post-inflationary evolution scenarios as defined in Eq.~(\ref{eq:def_R_RD-dS-RD}), for $\tau_{R_2}/\tau_{R_1}=1.9$ and $\tau_{R_2}/\tau_{R_1} = 1.3$, respectively. }
\label{fig:ratio-RD-dS-RD}
\end{figure}

$R$ depends only on $\tau_{R_1}$ and $\tau_{R_2}$. For deep IR modes, one can get $R \rightarrow 1$. While for deep UV we have $R \rightarrow \brr{2 - \tau_{R_2}/\tau_{R_1}}^2$. This is because when $k \rightarrow \infty$, the modes are always within the horizon, and therefore one can ignore the $a''/a$ term and $a \mathcal{E}$ in deep RD is determined by the initial condition which is proportional to $a_{\rm end}^{2} H_{\rm end}$. Fig.~\ref{fig:ratio-RD-dS-RD} shows the corresponding behavior of R function when taking different parameters. The left panel is for $\tau_{R_2}/\tau_{R_1} = 1.9$, as a consequence, for the deep UV mode the value of $R$ approaches 0.01. The blue curve refers to $\tau_{R_1}/|\tau_\star|=1/5$. In this case, at the beginning of the intermediate dS stage, only modes with $k > a_\star H_\star/5$ has re-entered horizon, and these modes will leave oscillatory imprints on $R$. The modes which have not re-entered the horizon will be out of the horizon until after the end of the intermediate dS stage. Therefore, $R$ is order one during the intermediate dS stage. The values of $\tau_{R_1}/|\tau_\star|$ for brown and red curves are 10 and 30, which is equivalent to moving the blue curve to low $k$ direction in correspondence. The right figure is under $\tau_{R_2}/\tau_{R_1} = 1.3$ in which the UV asymptotic value is enhanced. This is intuitive cause the longer the second dS inflation lasts for, the more suppressed for the deep UV modes. In next section we will show the signals under this RD-dS-RD evolution scenario after considering the explicit phase transition sources.

\section{Retarded Green's Function in inflating Universe}

\begin{figure}[t]
\centering
\includegraphics[height=5.1cm]{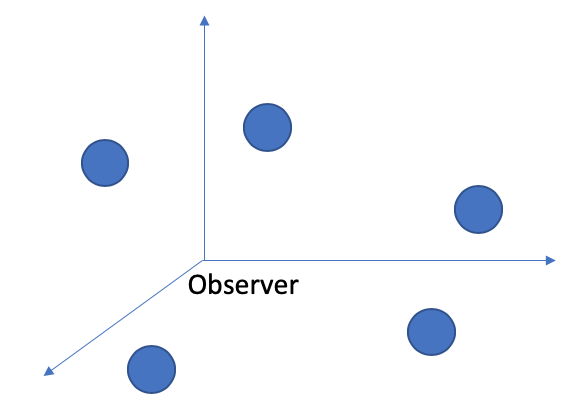}
\caption{Spatial distribution of the GW in de Sitter space at $|\tau| \gg |\tau_\star|$. }
\label{fig:balls}
\end{figure}

Here we present the results of the retarded Green's function in spatial coordinates. In de Sitter space, the spatial Green's function can be derived straightforwardly from the inverse Fourier transformation of Eq.~(\ref{eq:GIF}). Its Fourier transformation gives
 \begin{equation}\label{eq:greendS}
 \begin{split}
     G_{\rm dS}(\tau,\tau';\vec{x}-\vec{x}')
     &=\dfrac{H_{\rm inf} \tau}{4 \pi |\bx - \bx'|} \delta(\tau-\tau'-|\bx - \bx'|) + \dfrac{H_{\rm inf}}{4\pi\tau' }  \Theta(\tau-\tau'-|\bx-\bx'|) \ ,
\end{split}     
\end{equation}
where $\tau$ is the conformal time. There are two terms in $G_{\rm dS}$. The first term is proportional to $\delta(\tau-\tau' - |\bx - \bx'|)$. This term is similar to the retarded Green's function in Minkovski space describing the propagation of a massless scalar field produced at $\bx'$ at $\tau'$. The difference is that there is a red shift in the expanding universe described by the $\tau$ factor in front of the delta function. The second term in $G_{\rm dS}$ is new. Its spatial distribution is like a solid ball, with the field value evenly distributed inside. Specifically, at the IR boundary $\tau\rightarrow 0$, the delta function part vanishes due to the redshift. Therefore, in the case of first order phase transition during inflation, at the time $|\tau| \ll |\tau_\star|$ the spatial configuration of the GWs is like a bunch of solid balls scattered in the space with the comoving radius around $|\tau_\star|$, as shown in Fig.~\ref{fig:balls}. 

For $t^p$ inflation, the Green's function in coordinate space can be got from the inverse Fourier transformation of Eq.~(\ref{greenfunctions}). For finite $\tau$ the expression is involved due to the appearance of the special functions. For simplicity, we present here the Green's function in coordinate space in the limit $\tau\rightarrow 0$,
\begin{equation}
 \begin{split}
     G_{t^p}(\tau';\vec{x}-\vec{x}') &= \dfrac{(-\tau')^{-2} \Gamma\brr{\alpha}  }{2 a(\tau') \pi^{3/2} \Gamma(\alpha-1/2) } \brr{1- \dfrac{|\vec{x} - \vec{x}'|^2}{\tau'^2}}^{\alpha - 3/2} \Theta(-\tau'-|\vec{x}-\vec{x'}|)
\end{split}     
\end{equation}
One can see that the spatial distribution is still like a solid ball, but the distribution of the field inside the ball has a non-trivial distribution. In the limit of $p\rightarrow \infty$, the $t^p$ inflation approaches to de Sitter inflation and the $\alpha$ approaches to $3/2$, and we reproduce the result in Eq.~(\ref{eq:greendS}).


\bibliography{GW}
\bibliographystyle{JHEP}

\end{document}